\documentclass[preprint,12pt]{elsarticle}
\usepackage{graphicx}
\usepackage{alphabeta}
\usepackage{booktabs}
\usepackage{longtable}
\usepackage{ragged2e}
\usepackage{array}
\usepackage{pdflscape}
\usepackage{float}
\usepackage{hyperref}

\begin{document}
\begin{frontmatter}
     \title{Generative AI for Enzyme Design and Biocatalysis}
    \author[label1,label2]{Lasse Middendorf}
    \author[label1,label2,cor]{Noelia Ferruz}
    \cortext[cor]{Correspond to noelia.ferruz@crg.eu}
    
    \affiliation[label1]{organization={Centre For Genomic Regulation, the Barcelona Institute of Science and Technology},
            city={Dr Aiguader 88, Barcelona 08003},
            country={Spain}}
    \affiliation[label2]{organization={Universitat Pompeu Fabra},
            city={Barcelona},
            country={Spain}}

\begin{abstract}
Sparked by innovations in generative artificial intelligence (AI), the field of protein design has undergone a paradigm shift with an explosion of new models for optimizing existing enzymes or creating them from scratch. After more than one decade of low success rates for computationally designed enzymes, generative AI models are now frequently used for designing proficient enzymes. Here, we provide a comprehensive overview and classification of generative AI models for enzyme design, highlighting models with experimental validation relevant to real-world settings and outlining their respective limitations. We argue that generative AI models now have the maturity to create and optimize enzymes for industrial applications. Wider adoption of generative AI models with experimental feedback loops can speed up the development of biocatalysts and serve as a community assessment to inform the next generation of models. 
\end{abstract}

\begin{highlights}
\item Generative AI can create proficient enzymes for industrial biocatalysis
\item Language and diffusion models enable de novo protein design
\item Experimental feedback loops can accelerate enzyme engineering workflows and model development
\end{highlights}


\end{frontmatter}

\section{Introduction}
\label{Introduction}

Designing enzymes with tailored properties is an outstanding goal in biochemistry, with tremendous potential for biomanufacturing and synthetic biology \cite{bornscheuer2012thirdWaveBiocatalysis, erb2017syntheticMetabolism}. Most commonly, enzymes have been modified via random mutations in directed evolution or (semi-)rational design, taking into account the target enzyme's mechanism and structure \cite{arnold1998Directed_evolution,lutz2010semi_rational_engineering}. Although both approaches generated improved enzymes for a variety of reactions with examples of translation to industrial processes \cite{wu2021biocatalysisInIndustry1}, they face conceptual challenges that limit their success: Directed evolution relies on a (promiscuous) starting activity, and (semi-)rational design requires deep mechanistic knowledge of the property of interest. Moreover, both approaches greedily climb the fitness landscape towards the closest peak, with no guarantee of reaching the global maximum \cite{tokuriki2012diminishingReturns}.
\newline
\newline
An alternative to directed evolution and (semi-)rational approaches has been the bottom-up design of \textit{de novo} enzymes based on physicochemical principles. Here, the design process begins by arranging amino acid side-chains to stabilize the transition state of the reaction one wishes to catalyze. Then, a backbone that can harbour the active site design is found or created. Finally, the remaining protein sequence is generated to fold into the corresponding backbone \cite{chu2024sparksOfFunction}. \textit{De novo} enzymes, designed according to physicochemical principles, demonstrated feasibility but have low sucess rates and lacked the catalytic efficiencies needed for industrial applications without further optimization \cite{rothlisberger2008deNovoKempEliminase, jiang2008deNovoRetroAldolase, siegel2010deNovoDieselAlder, richter2012esteraseDesign}. The low catalytic efficiency is likely caused by reducing the enzyme solely to its active site, without considering contributions from electrostatic pre-organization and dynamics during backbone- and sequence-generation \cite{ruiz2024electrostatics, hammes2006Dynamcis}.
\newline
\newline
Nevertheless, recent advances in self-supervised deep learning have enabled the development of generative AI models for enzyme design \cite{romero2023exploringSequenceSpaceWithGenerativeModels}. By pre-training the models on sequences and structures across diverse families and folds, these models learn the inherent properties of functional proteins. Many different model architectures exist, and each has its own strengths and weaknesses. Here, we do not aim to give a technical overview of the different model architectures, or their downstream applications in machine learning-assisted directed evolution, as excellent reviews written for biologists already exist \cite{bepler2021learningTheProteinLanguage, romero2023exploringSequenceSpaceWithGenerativeModels, albanese2025computationalProteinDesign, yang2024MLDE_Review}. Instead, we highlight enzymes designed with the help of generative AI models, show which model architecture is suited to which design task, and discuss their respective limitations for real-world design campaigns.    

\section{Generative AI Models For Designing Sequence And Structure}
\label{Model classes}
\begin{figure}[H]
    \centering
    \includegraphics[width=\linewidth]{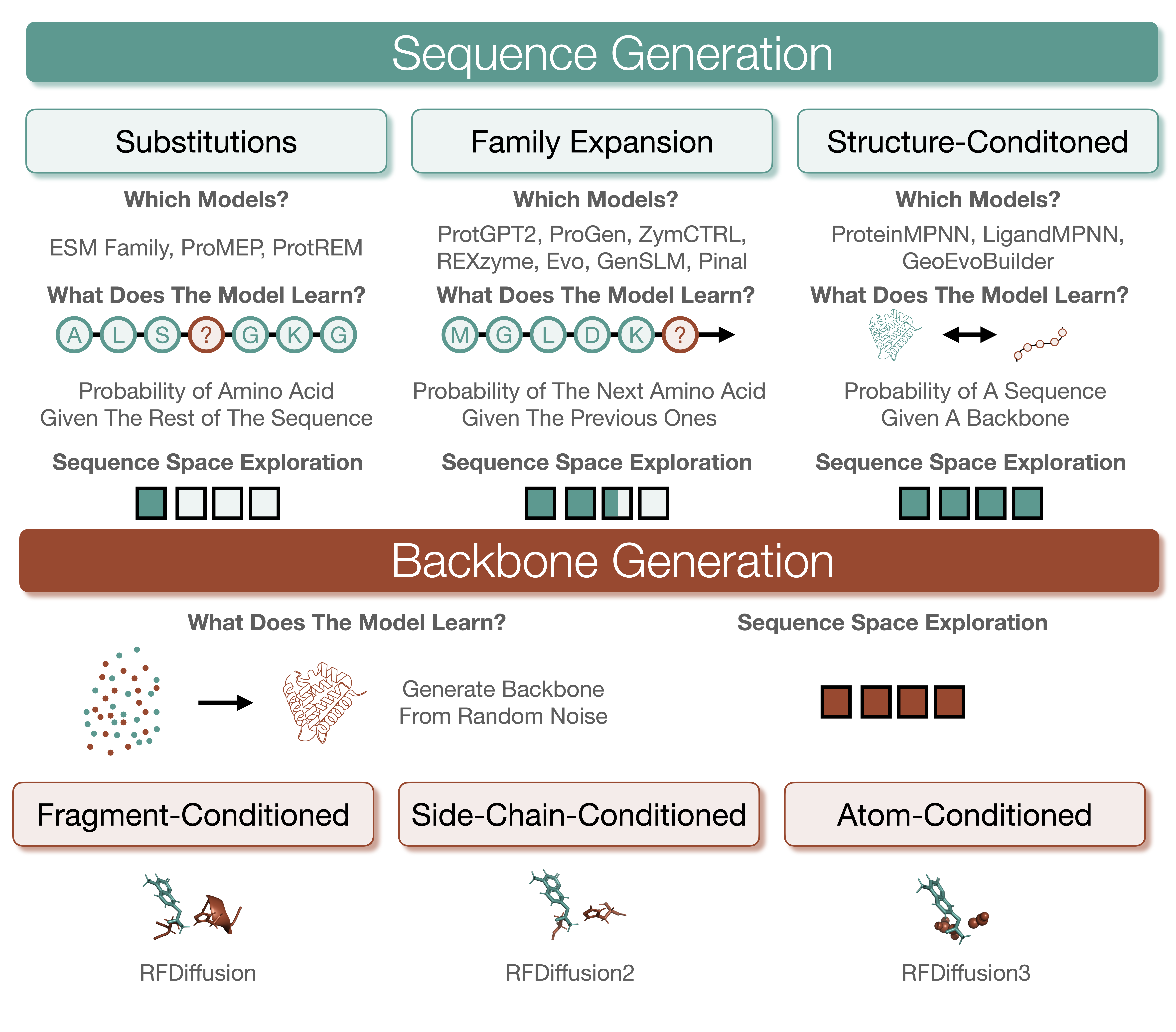}
    \caption{\textbf{Classes of Generative AI models For Enzyme Design} We classify generative AI models for enzyme design based on the modality they sample from. Sequence-generating models include substitution models to optimize existing enzymes, family expansion models to generate diverse novel sequences within a family, and structure-conditioned models that optimize sequences for fixed backbones. Backbone-generating models generate protein structures from noise. These are further distinguished by conditioning granularity, including fragment-conditioned, side-chain-conditioned, and atom-conditioned approaches for active site scaffolding}
    \label{fig1}
\end{figure}

We define generative AI models as any deep neural network that learns and samples complex, high-dimensional probability distributions underlying training data, in this case protein sequences, structures, and/or functions. In this review, we focus on generalist models that are trained across protein families. While examples of model architectures trained on the individual family level exist (e.g., variational autoencoders or generative adversarial networks), their performance needs to be evaluated on a case-by-case basis and require users to re-train them on their specific families of interest. We focus on models that do not necessarily require family-specific training, and can thus be used by non-expert users for real-world applications. We classify models according to their applications in enzyme design workflows (see Figure\ref{fig1}).   

\subsection{Sequence-Generating Models}
\label{Sequence-Generating Models}
Several models leverage the sequential nature of proteins to produce amino acid sequences tailored for different design tasks directly. Sequence-generating models can be applied to (1) introduce substitutions into existing sequences, (2) explore and expand enzyme families by proposing diverse yet plausible variants, or (3) enable (partial) redesign of sequences conditioned on protein structure (Fig.~\ref{fig1}).
\newline
In the first category, widely used approaches include encoder-only Transformer architectures trained with a masked language modelling (MLM) objective, i.e. predicting the identity of masked amino acids given the remainder of the sequence. Variants of these models incorporate additional context such as protein structure or multiple sequence alignments (MSAs), or combine multiple inputs \cite{bepler2021learningTheProteinLanguage, romero2023exploringSequenceSpaceWithGenerativeModels}. During inference, a full protein sequence (or sequence plus auxiliary context) is provided to the model, which returns position-wise log-probabilities (or log-likelihoods) over the 20 amino acids. These scores can be used to propose mutations (e.g., substitutions that increase model likelihood, or more generally that shift residues toward higher-probability alternatives under the model) and to produce input embeddings for downstream models. Selecting point mutations guided by this class of models can yield a higher density of improved variants than traditional random approaches such as error-prone PCR, as demonstrated by engineering the leaf-compost cutinase for higher activity towards PET guided by the MSA Transformer model \cite{maguire2024ESM_cutinase_petase}. Further, substitution selection can be refined with experimental data: adding a regression head trained on assay measurements on top of the MSA Transformer enabled guided engineering of leaf-compost cutinase variants with higher activity towards PET than those generated via random mutagenesis (1.62-fold improvement over wild type) \cite{maguire2024ESM_cutinase_petase}. This supports data-driven enzyme engineering loops, as demonstrated for the Halide Methyltransferase from \textit{Arabidopsis thaliana} and the Phytase from \textit{Yersinia mollaretii} \cite{singh2025ESM_atHMT_phytases}. In that workflow, substitution proposals informed by ESM2 \cite{lin2023ESM2} were introduced via site-directed mutagenesis, activities were measured, and the resulting data were used to train machine-learning models that guided subsequent rounds. After four rounds, the two enzymes showed approximately 16- and 26-fold increases in activity over wild type \cite{singh2025ESM_atHMT_phytases}. Collectively, these studies illustrate how probabilistic sequence models can be used to prioritize point mutations that improve properties when the model scores correlate with the property of interest. Similar strategies have also been applied to improve the precision and editing efficiency of genome editing tools \cite{perrotta2024ESM_deaminase_church, he2024ESM_glycosylase, wei2025ProMEP_Cas9}.
\newline
Many recent enzyme-engineering studies have used sequence models from the ESM family (see Table~\ref{table1}) \cite{rives2021ESM1b, meier2021ESM1v, rao2021MSATransformer, lin2023ESM2}. At the same time, the most suitable model can depend on the target property, protein family, and available conditioning information (e.g., MSA depth or structural context). The ProteinGym benchmark evaluates how well model-derived amino acid probabilities correlate with experimentally measured effects of point mutations across phenotypes including activity, binding, expression, organismal fitness, and stability \cite{notin2023proteingym}. Its leaderboard therefore provides a useful reference for selecting models for substitution-guided design.
\newline

While masked-language models are typically used to score and prioritize a relatively small number of substitutions in a given scaffold, autoregressive sequence-generating models can propose entirely new sequences and thereby explore a broader region of sequence space\cite{ferruz2022protgpt2, madani2023Progen, nijkamp2023progen2, munsamy2024ZymCTRL, zvyagin2023GenSLM, nguyen2024Evo, dai2024Pinal}. These models are trained with a causal language modelling (CLM) objective, i.e. to predict the next amino a given the previous ones and are often trained with a decoder-only transformer architecture \cite{bepler2021learningTheProteinLanguage, romero2023exploringSequenceSpaceWithGenerativeModels}. Enzymes generated with family-expansion models have shown as little as 35\% sequence identity to any other known enzyme in the respective family (Table \ref{table1}) \cite{romero2024ProtGPT_ZymCTRL_TIM}.
While most of the design campaigns did not set out to generate enzymes with improved properties for biocatalysis (see Table\ref{table1}), enzymes with higher activity or stability relative to a natural reference were found in multiple cases \cite{munsamy2024ZymCTRL, armer2025AlignFoundation_AlphaAmylases, dai2024Pinal, ivanvcic2025Progen_Piggybac}. A retrospective analysis of designed alpha amylases revealed that the family expansion model used for their design produces sequences that recapitulate the properties of the sequences in its training data \cite{stocco2024ProtRL}. The improved properties of enzymes designed with family-expansion models are therefore likely due to that they are already present in the sequences the models were trained on. This raises the question of the current applicability of family expansion models for designing enzymes with desired properties for biocatalysis that are not found in natural sequences (see Section \ref{limitations}) and has prompted a wealth of recent work on the guiding of generative models \cite{stocco2025Steering_review}.
\newline
\newline

While family-expansion models can already explore large parts of the accessible sequence space of entire enzyme families, structure-conditioned models can generate even more distant sequences when prompted with certain structural scaffolds \ref{fig1}. During their training, they explicitly learn the relationship between sequences and structures \cite{dauparas2022ProteinMPNN, dauparas2025LigandMPNN, liu2025geoevobuilder}, enabling them also to generate sequences for protein structures that do not occur in nature \cite{dauparas2022ProteinMPNN}. Users of structure-conditioned generative models can fix specified residues in their input structure, enabling sequence generation from point mutations to full-length sequences. However, currently, full-length sequence generation has been applied only to \textit{de novo} generated backbones as input structures (see Section \ref{Backbone-Generating Models}). When applied to redesigning natural enzymes, the most conserved residues of the enzyme family and the active site are often fixed and not redesigned \cite{sumida2024ProteinMPNN_TEV, king2025ProteinMPNN_aKG}. The two most popular structure-conditioned generative models are ProteinMPNN and its further development, LigandMPNN, which takes interactions with non-protein molecules into account \cite{dauparas2022ProteinMPNN, dauparas2025LigandMPNN}. Both have extensive applications in enzyme design for enhancing stability and activity.

One of the first enzymes redesigned with ProteinMPNN was the TEV protease \cite{sumida2024ProteinMPNN_TEV}. Sequences were designed with a variable fraction of fixed residues, and the most active designs were found when 50\% of the most conserved residues were fixed. The best-performing design displayed a 26-fold increase in activity over the wild type and a remarkable 40°C increase in the melting temperature \cite{sumida2024ProteinMPNN_TEV}. Applying ProteinMPNN for sequence design generally produces sequences with high stability \cite{dauparas2022ProteinMPNN}. Even if increasing stability is not the main objective of the design campaign, increased stability gained through sequence design with ProteinMPNN can facilitate the introduction of activity-enhancing mutations, which are often destabilizing \cite{bloom2006protein_stabillity_promotes_evolvability, tokuriki2008stabillity_function_tradeoff}. Indeed, the prior stabilization of the \textalpha{}-ketoglutarate-dependent oxygenase tP4H via sequence design with ProteinMPNN allowed for the subsequent introduction of activity-enhancing mutations towards a promiscuous substrate. Directed evolution of the stabilized variant led to an 80-fold increase in the activity of tP4H, while only a 6-fold improvement was observed in the wild type background \cite{king2025ProteinMPNN_tP4H}. To leverage the benefits of stabilizing mutations when increasing activity, ProteinMPNN was included in the AI.zyme pipeline, which combines ProteinMPNN, structure prediction, and physicochemical methods for the computational evolution of optimized enzymes \cite{merlicek2025aizymes}. This pipeline was used to generate a ketosteroid isomerase with 7.7-fold higher Kemp eliminase activity by testing 7 designs with 45\% to 55\% sequence identity to the starting sequence \cite{merlicek2025aizymes}. While the previous examples relied on directed evolution or phiscochemical-based computational methods fo introduce mutations that affect promiscuous activity, ProteinMPNN and LigandMPNN can also be used directly to modify ligand-interacting residues \cite{sun2025cmdmpnn, hou2025denovo_porphyrin_containing_proteins, huang2025denovo_GeH_insertion}. ProteinMPNN was combined with MD simulations of the protein complex with a promiscuous ligand to redesign active-site residues and identify variants with binding modes and distances similar to those of the natural substrate \cite{sun2025cmdmpnn}. This broadened the substrate scope of the \textit{A. thaliana} glycosyltransferase for phenolic pollutants \cite{sun2025cmdmpnn}. However, as ProteinMPNN has not been trained on structures containing non-protein atoms, its application to predicting interface amino acids is challenging \cite{dauparas2025LigandMPNN} and requires careful consideration. In contrast, LigandMPNN \cite{dauparas2025LigandMPNN}, was extended to learn from structures including non-protein atoms, and is suitable for redesigning residues that interact with ligands. LigandMPNN was applied to improve the enantioselectivity of cyclopropanation, silylation, and organogermane synthesis using a previously \textit{de novo} designed four-helix bundle \cite{hou2025denovo_porphyrin_containing_proteins, huang2025denovo_GeH_insertion}. Here, LigandMPNN was used to design enzymes that  stabilize only the transition state of the preferred enantiomer, while introducing clashes with others. These examples highlight the potential of structure-conditioned sequence-generating models for designing enzymes catalyzing new-to-nature reactions, when paired with \textit{de novo} designed backbones capable of harbouring tailored active sites. 

\begin{figure}[H]
    \centering
    \includegraphics[width=\linewidth]{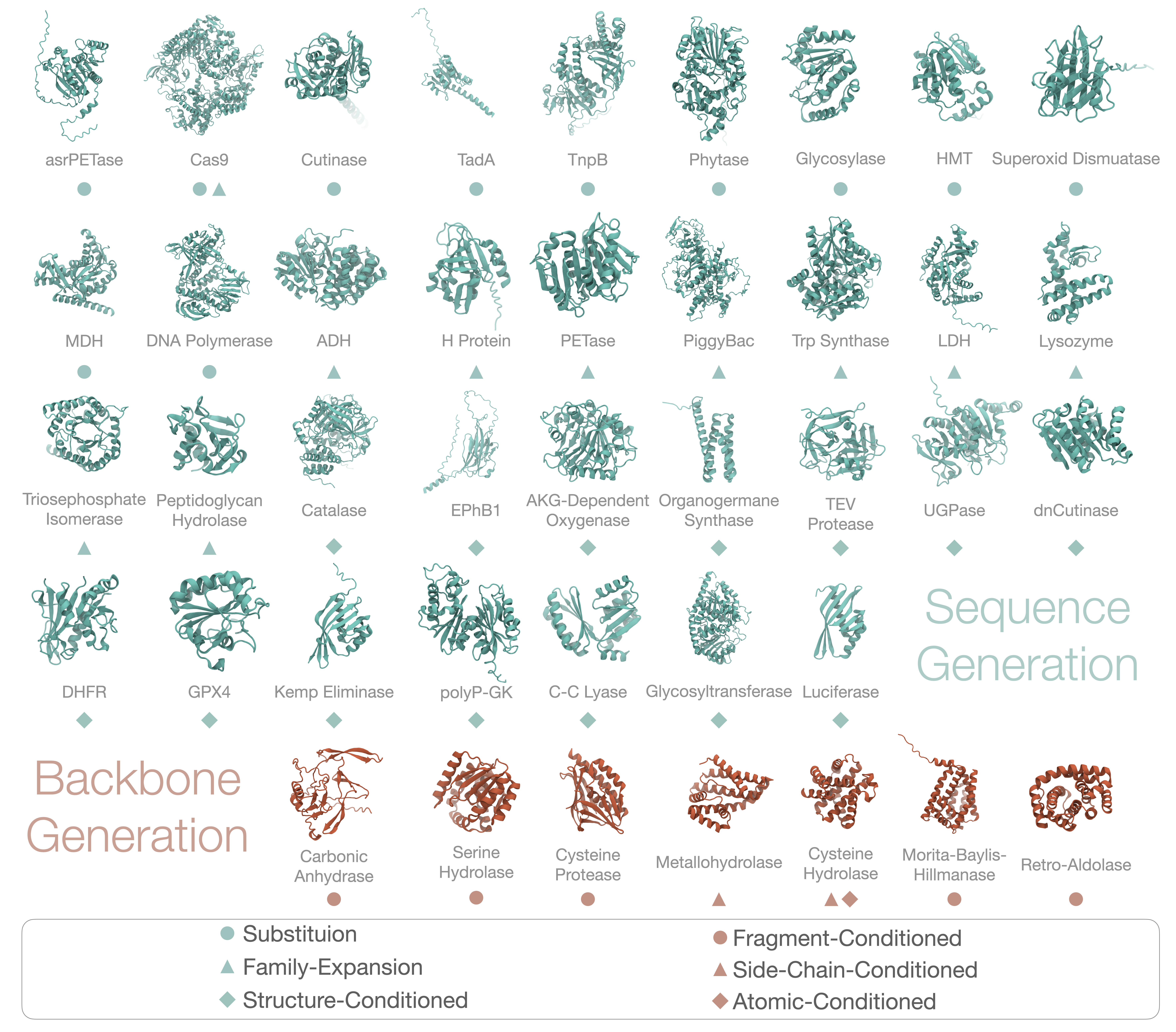}
    \caption{\textbf{Enzymes Designed With Generative AI}. Structures of the best performing best performing enzymes designed with generative AI models. The structures were retrieved from the original publication, the AlphaFold database (for designs only containing few substitutions), or predicted with ESMFold. In cases where multiple designs for the same enzyme exist, we display the most active one. Abbreviations: Alcohol Dehydrogenase (ADH), Dihydrofolate Reductase (DHFR), Halide Methyltransferase (HMT), Lactate Dehydrogenase (LDH), Malate Dehydrogenase (MDH)}
    \label{fig2}
\end{figure}

\subsection{Backbone-Generating Models}
\label{Backbone-Generating Models}
Backbone-generating models allow the design of structures harboring defined active sites capable of stabilizing transition states for variable reactions. During training, they learn to recover protein structures from random noise and thus approximate the inherent probability of natural protein structures. They substantially contributed to the acceleration of \textit{de novo} enzyme design by almost doubling the number of successful design \textit{de novo} campaings in the last two years alone. Some of the most popular backbone-generating models are the RFDiffusion family \cite{watson2023RFdiffusion, ahern2025RFDiffusion2, butcher2025RFdiffusion3}, each allowing a different granularity level of conditioning of the backbone generation on an input. 

The first enzyme designed by conditioning the backbone-generating model RFDiffusion on an active site was a serine hydrolase, which catalyzes ester bond hydrolysis in a two-step reaction \cite{lauko2025serin_hydrolase}. RFDiffusion was conditioned on two tripeptides containing catalytic serine and histidine, respectively. The geometry of the theozyme was optimized with respect to the transition state structure using density functional theory (DFT). The generated backbones were screened with Rosetta for their ability to accommodate aspartate or glutamate in the active site to complement the catalytic triads of serine hydrolases. Finally, sequences for the respective backbones were generated via iterative applications of LigandMPNN and relaxation, followed by the final structure prediction with AlphaFold2. The best-performing design had a  catalytic efficiency of 2.2 x 10\textsuperscript{5} M\textsuperscript{-1} s\textsuperscript{-1}, which is slower than natural serine hydrolases with their native substrate but reaches the median efficiencies of natural enzymes \cite{lauko2025serin_hydrolase, bar2011moderately}. RFDiffusion was further used within the Riff-Diff pipeline to design \textit{de novo} retro-aldolases and Morita–Baylis–Hillmanases with activities comparable to those of previously engineered enzymes. Similar to the design of the \textit{de novo} serine hydrolases, the Riff-Diff pipeline starts from a catalytic array known to catalyze the reaction, with the catalytic residues placed within an \textalpha{}-helix. After building a motif library of suitable backbone fragments to host the desired active-site rotamers, the rest of the enzyme backbone is generated with RFDiffusion, and sequences are generated similarly to those for the \textit{de novo} serine hydrolases \cite{braun2025RiffDiff}. 
The introduction of RFdiffusion2 \cite{ahern2025RFDiffusion2} addressed many of the problems that were present in earlier models. RFDiffusion2 bypasses the search for backbone fragments capable of hosting the active site rotamers and does not require an \textit{a priori} assignment of the active site residues' positions in the sequence. By providing only the coordinates of the active site residues and the ligand, RFDiffusion2 was capable of designing backbones for retroaldolases, cysteine and zinc hydrolases, and cysteine proteases \cite{ahern2025RFDiffusion2, choi2025RFD2_cystein_protease}. While most of these \textit{de novo} enzymes have low catalytic efficiencies, some of the designed zinc hydrolases reach activity levels similar to natural ones with similar substrates. A retrospective analysis of these designs showed a very similar and confident placement of the active site residues in the designed enzymes and the calculated theozyme \cite{kim2025RFd2_metallohydrolases}. However, a more in-depth exploration of predictors for high enzymatic activity will be necessary, as highly accurate placement of residues directly interacting with the substrate is not always sufficient to rank designs by activity levels \cite{braun2025RiffDiff, lauko2025serin_hydrolase}. The current generation of backbone-generating models from the RFDiffusion family \cite{ahern2025RFDiffusion2, butcher2025RFdiffusion3} or Proteus2 \cite{qu2026proteus2_metalloprotease}, paired with structure-conditioned sequence-generating models, shifts the design problem towards creating potent active sites. The designs highlighted above show that our capabilities already moved beyond reactions like the Kemp elimination towards multi-step reactions and non-activated substrates \cite{lauko2025serin_hydrolase, choi2025RFD2_cystein_protease, qu2026proteus2_metalloprotease}.

\section{Limitations of Current Model Types}
\label{limitations}
The design campaigns outlined in the previous sections highlight that generative AI models can already be used to optimize biocatalysts for multiple properties relevant for industrial applications. However, specific conceptual challenges remain in designing proteins with properties fundamentally different from those of their natural counterparts. The training data of all models is inherently shaped by evolutionary principles that prioritize biological fitness over performance in industrial settings. Additionally, the training objective of substitution and family expansion models appears to reward features that are not directly related to protein fitness \cite{hou2025understanding_plm_scaling_fitness_prediction}. Nevertheless, post-training and steering techniques with experimental data are rapidly developing approaches to align the generative AI models towards the property of interest to generate improved sequences \cite{stocco2024ProtRL, stocco2025Steering_review, widatalla2024ProteinDPO, blalock2025RLXF}. Structure-conditioned and backbone-generating models have shown a capacity for \textit{de novo} enzyme design that generalizes well beyond the evolution of natural proteins (see Table\ref{table1}). But frameworks have yet to integrate all the physico-chemical factors that govern catalysis. For example, all \textit{de novo} designed enzymes backbones discussed before mostly neglected second-shell interactions or contributions from the overall fold, which could improve activity \cite{ruiz2024electrostatics}.
Finally, no "end-to-end" solution currently exists for designing enzymes with new-to-nature activities, as these campaigns still require significant manual intervention and a non-trivial chemical hypothesis. No models can yet autonomously deduce the chemical requirements for a novel reaction as current generative architectures largely learn statistical regularities from sequence and structure data. Future approaches will most likely require the explicit incorporation of physico-chemical priors in the models.

\section{Conclusion}
\label{Conclusion}
Generative AI models for protein design are developing rapidly and are mostly integrating improvements from natural language modeling and image generation. The current generation of sequence-generating models attributes their improvements to increased parameter counts, which have led to considerable gains in natural language modeling but already show diminishing returns on benchmarking sets \cite{bhatnagar2025Progen3, brixi2025evo2, hou2025understanding_plm_scaling_fitness_prediction} and are rarely evaluated in task-relevant real-world enzyme design campaigns. For meaningful improvements in the next generation of generative AI models for enzyme design, it will be necessary to test them under realistic conditions and evaluate where they excel and where they fail. The examples of generative AI models highlighted here show their maturity for real-world applications and motivate further exploration, especially where traditional methods fail. Especially when paired with automated testing \cite{singh2025ESM_atHMT_phytases}, integration of experimental data \cite{stocco2024ProtRL, widatalla2024ProteinDPO, blalock2025RLXF}, and potentially explainable AI, generative AI models can be powerful tools for efficiently creating biocatalysts that meet industrial needs and for better understanding the fundamental mechanisms underlying enzymatic catalysis.

\begin{landscape}

\thispagestyle{empty}

\footnotesize
\renewcommand{\arraystretch}{1.25}

\setlength{\LTpre}{8mm}
\setlength{\LTpost}{0mm}

\begin{longtable}{
    >{\raggedright\arraybackslash}p{3.0cm}
    >{\raggedright\arraybackslash}p{3.0cm}
    >{\raggedright\arraybackslash}p{1.5cm}
    >{\raggedright\arraybackslash}p{3.0cm}
    >{\raggedright\arraybackslash}p{2.7cm}
    >{\raggedright\arraybackslash}p{2.0cm}
    >{\raggedright\arraybackslash}p{2.5cm}
    p{1.0cm}
}
\caption{\textbf{Summary of Engineered Enzymes and Computational Models} The sequence identity of the best performing design was calculated from the closest sequence in UniRef90 with the exception for enzymes where the best design was reported to have single substitutions only. Alignment was performed with MMSeqs easy-search with default settings.} \label{table1} \\
\toprule
\textbf{Enzyme} & \textbf{Model(s)} & \textbf{Pipeline} & \textbf{Objective} & \textbf{Model Type} & \textbf{Seq. Id. [\%]} & \textbf{Improvement / Activity} & \textbf{Ref.} \\
\midrule
\endfirsthead

\caption[]{\textbf{Summary of Engineered Enzymes and Computational Models} (Continued)} \\
\toprule
\textbf{Enzyme} & \textbf{Model(s)} & \textbf{Pipeline} & \textbf{Objective} & \textbf{Model Type} & \textbf{Seq. Id. [\%]} & \textbf{Improvement / Activity} & \textbf{Ref.} \\
\midrule
\endhead

\midrule
\multicolumn{8}{r}{\textit{Continued on next page}} \\
\midrule
\endfoot

\bottomrule
\endlastfoot

Alcohol Dehydrogenase & Pinal & -- & Redesign & Family Expansion & 49 &  Not Reported & \cite{dai2024Pinal} \\

$\alpha$-Ketogluterate-dependent Oxygenase & ProteinMPNN & -- & Stability & Structure-Conditioned & 84 & 80-fold &  \cite{king2025ProteinMPNN_aKG} \\

asrPETase & ESM1v & -- & Activity & Substitution & Not Determined & 6.1-fold & \cite{song2025ESM_ASR_Petase} \\

C-C lyase & ProteinMPNN & -- & Oligomeric State & Structure-Conditioned & 49 & 5 M\textsuperscript{-1} s\textsuperscript{-1} & \cite{egea2025AI_lanthanide_lyase}\\

Carbonic Anhydrases & RFDiffusion + ProteinMPNN & GRACE & De Novo Design & Backbone-Generating, Structure-Conditioned & 46 & Not Reported & \cite{hu2024grace_denovo_Carbonic_anhydrase}\\

Cas9 & ProMEP, Progen2  & -- & Activity & Substitution, Family Expansion & Not Determined, 71 & 2-3-fold; No Improvement &\cite{wei2025ProMEP_Cas9, ruffolo2025OpenCRISPR}\\

Catalase & ProteinMPNN & -- & Activity, Stability & Structure-Conditioned & 99 & 1.9-fold & \cite{xu2025ProteinMPNN_catalase}\\

Copper Superoxide Dismutase & MSATransformer & -- & Redesign & Substitution & 72 & Not Reported & \cite{johnson2025protein_scoring}\\

Cutinase & MSATransformer, ProteinMPNN & -- & Activity, Expression & Substitution, Structure-Conditioned & 38 & 1.62-fold; No Improvement & \cite{maguire2024ESM_cutinase_petase, ding2025denovo_cutinase}\\

Cyclopropane Synthases & LigandMPNN & -- & Substrate Scope & Structure-Conditioned & 46 & 95-fold & \cite{hou2025denovo_porphyrin_containing_proteins}\\

Cystein Hydrolase & RFDiffusion2 + ProteinMPNN; RFDiffusion3 + LigandMPNN & -- & De Novo Design & Structure-Conditioned, Backbone-Generating & No hits &  12 M\textsuperscript{-1} s\textsuperscript{-1};  3.5 x 10\textsuperscript{4} M\textsuperscript{-1} s\textsuperscript{-1} & \cite{choi2025RFD2_cystein_protease,butcher2025RFdiffusion3}\\

Dihydrofolate Reductase & GeoEvoBuilder & -- & Activity, Stability & Structure-Conditioned & 84 & 21.05-fold & \cite{liu2025geoevobuilder}\\

EPhB1 Tyrosine Kinase & ProteinMPNN, Frame2Seq & -- & Redesign & Structure-Conditioned & 96 & 1.3-fold & \cite{seki2025Ephb1_proteinMPNN_frame2seq}\\

Glutathione Peroxidase 4 & GeoEvoBuilder & -- & Stability, Activity & Structure-Conditioned & 88 & 8.49-fold & \cite{liu2025geoevobuilder}\\

Glycosyltransferase & ProteinMPNN & CMDmpnn & Substrate Scope & Structure-Conditioned & 99 & $\le$5-fold & \cite{sun2025cmdmpnn}\\

H-Protein & Pinal & -- & Redesign & Family Expansion & 52 & 1.7-fold & \cite{dai2024Pinal}\\

Halide Methyltransferase & ESM2 & -- & Activity & Substitution & Not Determined & 16-fold & \cite{singh2025ESM_atHMT_phytases}\\

Kemp Eliminase & ProteinMPNN & AI.zymes & Activity & Structure-Conditioned & 52 & 7.7-fold & \cite{merlicek2025aizymes}\\

Lactate Dehydrogenase & ZymCTRL & -- & Redesign & Family Expansion & 87 & 3-fold & \cite{munsamy2024ZymCTRL}\\

Luciferase & ProteinMPNN & -- & De Novo Design & Structure-Conditioned & No hit & 10\textsuperscript{6} M\textsuperscript{-1} s\textsuperscript{-1} & \cite{yeh2023denovo_luciferase}\\

Lysozyme & ProGen & -- & Redesign & Sequence, Family Expansion & 84 & No Improvement & \cite{madani2023Progen}\\

Malate Dehydrogenase & MSATransformer & -- & Redesign & Substitution & 78 & Not Reported & \cite{johnson2025protein_scoring}\\

Morita-Baylis-Hillmanase & RFDiffusion + LigandMPNN & Riff-Diff & De Novo Design & Structure-Conditioned, Backbone-Generating & No hit & 7.7 x 10\textsuperscript{-3}  min\textsuperscript{-1} & \cite{braun2025RiffDiff}\\

Organogermane Synthase & LigandMPNN & -- & Substrate Scope & Structure-Conditioned & 47 & Not Reported & \cite{huang2025denovo_GeH_insertion}\\

Peptidoglycan Hydrolase & PROPEND & -- & Redesign & Family Expansion & -- & Not Reported & \cite{wang2024Propend}\\

PETase & Pinal & -- & Redesign & Family Expansion & 34 & No improvement &  \cite{dai2024Pinal}\\

phi27 DNA Polymerase & ProtREM & -- & Activity, Stability & Substitution & Not Determined & 6.5-fold & \cite{tan2024ProtREM}\\

Phytase & ESM2 & -- & Activity & Substitution & Not Determined & 26-fold & \cite{singh2025ESM_atHMT_phytases}\\

PiggyBac Transposase & ProGen2 & -- & Activity & Family Expansion & 95 & 2-fold & \cite{ivanvcic2025Progen_Piggybac}\\

Polyphosphate Glucokinase & ProteinMPNN & CoSaNN & Stability & Structure-Conditioned & 53 & Not Reported & \cite{zimmerman2024CoSann_polyphosphate_glucokinase}\\

Retro-Aldolases & RFDiffusion + LigandMPNN & Riff-Diff & De Novo Design & Backbone-Generating, Structure-Conditioned & No hit & 290 M\textsuperscript{-1} s\textsuperscript{-1} & \cite{braun2025RiffDiff}\\

Serine Hydrolase & RFDiffusion + LigandMPNN & -- & De Novo Design & Backbone-Generating, Structure-Conditioned & No hit & 2.2 x 10\textsuperscript{5} M\textsuperscript{-1} s\textsuperscript{-1} & \cite{lauko2025serin_hydrolase}\\

TadA Deaminase & ESM1b, ESM1v, ProMEP & -- & Activity & Substitution & Not Determined & $\le$1.25-fold & \cite{perrotta2024ESM_deaminase_church}\\

TEV Protease & ProteinMPNN & -- & Activity, Stability & Structure-Conditioned & 72 & 26-fold & \cite{sumida2024ProteinMPNN_TEV}\\

TnpB Transposase & ProMEP & -- & Activity & Substitution & Not Determined & 3-fold & \cite{cheng2024ProMEP_tnbp}\\

Triosephosphate Isomerase & ProtGPT2, ZymCTRL & -- & Redesign & Family Expansion & 35 & 7.3 x 10\textsuperscript{6} M\textsuperscript{-1} s\textsuperscript{-1} & \cite{romero2024ProtGPT_ZymCTRL_TIM}\\

Tryptophan Synthase & GenSLM & -- & Stability, Substrate Scope & Family Expansion & 79 & 5-fold & \cite{lambert2026tryptophane_synthase}\\

UDP-Glucose Pyrophosphorylase & ProteinMPNN & -- & Stability & Structure-Conditioned & 99 & 500-fold & \cite{li2024UDP_glucose_pyrophosphorylase}\\

Uracil-N Glycosylase & ESM1v, ESM2 & -- & Activity & Substitution & Not Determined & 2-fold & \cite{he2024ESM_glycosylase}\\

Zinc Metallohydrolase & RFDiffusion2 + LigandMPNN, Proteus2 & -- & De Novo Design & Structure-Conditioned, Backbone-Generating & No hit & 1.6 x 10\textsuperscript{4} M\textsuperscript{-1} s\textsuperscript{-1} & \cite{ahern2025RFDiffusion2, kim2025RFd2_metallohydrolases}\\

\end{longtable}
\end{landscape}

\section{Acknowledgments}
\label{Acknowledgments}
This publication and other research outcomes are supported by the predoctoral program AGAUR-FI ajuts (2025 FI-3 00065) Joan Oró, of the Department of Research and Universities of the Generalitat of Catalonia, as well as the European Social Plus Fund. NF acknowledges support from the Ramón y Cajal Fellowship(RYC2021-034367-I) funded by MICIU/AEI/10.13039/
501100011033 and the European Union NextGeneration EU/PRTR, and from the European Union’s Horizon Europe program under grant agreement No 101165231. We acknowledge support of the Spanish Ministry of Science andInnovation through the Centro de Excelencia Severo Ochoa (CEX2020-001049-S, MCIN/AEI /10.13039/501100011033), and the Generalitat de Catalunya through the CERCA programme. This work is supported by the grant ATHENA (ERC-ST-2024, Grant agreement 101165231). Funded by the European Union. Views and opinions expressed are however those of the author(s) only and do not necessarily reflect those of the European Union. Neither the European Union nor the granting authority can be held responsible for them.

\bibliographystyle{elsarticle-num-names} 
\bibliography{cocb}

\begin{thebibliography}{80}
\expandafter\ifx\csname natexlab\endcsname\relax\def\natexlab#1{#1}\fi
\providecommand{\url}[1]{\texttt{#1}}
\providecommand{\href}[2]{#2}
\providecommand{\path}[1]{#1}
\providecommand{\DOIprefix}{doi:}
\providecommand{\ArXivprefix}{arXiv:}
\providecommand{\URLprefix}{URL: }
\providecommand{\Pubmedprefix}{pmid:}
\providecommand{\doi}[1]{\href{http://dx.doi.org/#1}{\path{#1}}}
\providecommand{\Pubmed}[1]{\href{pmid:#1}{\path{#1}}}
\providecommand{\bibinfo}[2]{#2}
\ifx\xfnm\relax \def\xfnm[#1]{\unskip,\space#1}\fi
%Type = Article
\bibitem[{Bornscheuer et~al.(2012)Bornscheuer, Huisman, Kazlauskas, Lutz,
  Moore, and Robins}]{bornscheuer2012thirdWaveBiocatalysis}
\bibinfo{author}{U.~T. Bornscheuer}, \bibinfo{author}{G.~Huisman},
  \bibinfo{author}{R.~Kazlauskas}, \bibinfo{author}{S.~Lutz},
  \bibinfo{author}{J.~Moore}, \bibinfo{author}{K.~Robins},
\newblock \bibinfo{title}{Engineering the third wave of biocatalysis},
\newblock \bibinfo{journal}{Nature} \bibinfo{volume}{485}
  (\bibinfo{year}{2012}) \bibinfo{pages}{185--194}.
%Type = Article
\bibitem[{Erb et~al.(2017)Erb, Jones, and
  Bar-Even}]{erb2017syntheticMetabolism}
\bibinfo{author}{T.~J. Erb}, \bibinfo{author}{P.~R. Jones},
  \bibinfo{author}{A.~Bar-Even},
\newblock \bibinfo{title}{Synthetic metabolism: metabolic engineering meets
  enzyme design},
\newblock \bibinfo{journal}{Current opinion in chemical biology}
  \bibinfo{volume}{37} (\bibinfo{year}{2017}) \bibinfo{pages}{56--62}.
%Type = Article
\bibitem[{Arnold(1998)}]{arnold1998Directed_evolution}
\bibinfo{author}{F.~H. Arnold},
\newblock \bibinfo{title}{Design by directed evolution},
\newblock \bibinfo{journal}{Accounts of chemical research} \bibinfo{volume}{31}
  (\bibinfo{year}{1998}) \bibinfo{pages}{125--131}.
%Type = Article
\bibitem[{Lutz(2010)}]{lutz2010semi_rational_engineering}
\bibinfo{author}{S.~Lutz},
\newblock \bibinfo{title}{Beyond directed evolution—semi-rational protein
  engineering and design},
\newblock \bibinfo{journal}{Current opinion in biotechnology}
  \bibinfo{volume}{21} (\bibinfo{year}{2010}) \bibinfo{pages}{734--743}.
%Type = Article
\bibitem[{Wu et~al.(2021)Wu, Snajdrova, Moore, Baldenius, and
  Bornscheuer}]{wu2021biocatalysisInIndustry1}
\bibinfo{author}{S.~Wu}, \bibinfo{author}{R.~Snajdrova}, \bibinfo{author}{J.~C.
  Moore}, \bibinfo{author}{K.~Baldenius}, \bibinfo{author}{U.~T. Bornscheuer},
\newblock \bibinfo{title}{Biocatalysis: enzymatic synthesis for industrial
  applications},
\newblock \bibinfo{journal}{Angewandte Chemie International Edition}
  \bibinfo{volume}{60} (\bibinfo{year}{2021}) \bibinfo{pages}{88--119}.
%Type = Article
\bibitem[{Tokuriki et~al.(2012)Tokuriki, Jackson, Afriat-Jurnou, Wyganowski,
  Tang, and Tawfik}]{tokuriki2012diminishingReturns}
\bibinfo{author}{N.~Tokuriki}, \bibinfo{author}{C.~J. Jackson},
  \bibinfo{author}{L.~Afriat-Jurnou}, \bibinfo{author}{K.~T. Wyganowski},
  \bibinfo{author}{R.~Tang}, \bibinfo{author}{D.~S. Tawfik},
\newblock \bibinfo{title}{Diminishing returns and tradeoffs constrain the
  laboratory optimization of an enzyme},
\newblock \bibinfo{journal}{Nature Communications} \bibinfo{volume}{3}
  (\bibinfo{year}{2012}) \bibinfo{pages}{1257}.
%Type = Article
\bibitem[{Chu et~al.(2024)Chu, Lu, and Huang}]{chu2024sparksOfFunction}
\bibinfo{author}{A.~E. Chu}, \bibinfo{author}{T.~Lu}, \bibinfo{author}{P.-S.
  Huang},
\newblock \bibinfo{title}{Sparks of function by de novo protein design},
\newblock \bibinfo{journal}{Nature biotechnology} \bibinfo{volume}{42}
  (\bibinfo{year}{2024}) \bibinfo{pages}{203--215}.
%Type = Article
\bibitem[{R{\"o}thlisberger et~al.(2008)R{\"o}thlisberger, Khersonsky,
  Wollacott, Jiang, DeChancie, Betker, Gallaher, Althoff, Zanghellini, Dym
  et~al.}]{rothlisberger2008deNovoKempEliminase}
\bibinfo{author}{D.~R{\"o}thlisberger}, \bibinfo{author}{O.~Khersonsky},
  \bibinfo{author}{A.~M. Wollacott}, \bibinfo{author}{L.~Jiang},
  \bibinfo{author}{J.~DeChancie}, \bibinfo{author}{J.~Betker},
  \bibinfo{author}{J.~L. Gallaher}, \bibinfo{author}{E.~A. Althoff},
  \bibinfo{author}{A.~Zanghellini}, \bibinfo{author}{O.~Dym}, et~al.,
\newblock \bibinfo{title}{Kemp elimination catalysts by computational enzyme
  design},
\newblock \bibinfo{journal}{Nature} \bibinfo{volume}{453}
  (\bibinfo{year}{2008}) \bibinfo{pages}{190--195}.
%Type = Article
\bibitem[{Jiang et~al.(2008)Jiang, Althoff, Clemente, Doyle, Rothlisberger,
  Zanghellini, Gallaher, Betker, Tanaka, Barbas~III
  et~al.}]{jiang2008deNovoRetroAldolase}
\bibinfo{author}{L.~Jiang}, \bibinfo{author}{E.~A. Althoff},
  \bibinfo{author}{F.~R. Clemente}, \bibinfo{author}{L.~Doyle},
  \bibinfo{author}{D.~Rothlisberger}, \bibinfo{author}{A.~Zanghellini},
  \bibinfo{author}{J.~L. Gallaher}, \bibinfo{author}{J.~L. Betker},
  \bibinfo{author}{F.~Tanaka}, \bibinfo{author}{C.~F. Barbas~III}, et~al.,
\newblock \bibinfo{title}{De novo computational design of retro-aldol enzymes},
\newblock \bibinfo{journal}{science} \bibinfo{volume}{319}
  (\bibinfo{year}{2008}) \bibinfo{pages}{1387--1391}.
%Type = Article
\bibitem[{Siegel et~al.(2010)Siegel, Zanghellini, Lovick, Kiss, Lambert,
  St.~Clair, Gallaher, Hilvert, Gelb, Stoddard
  et~al.}]{siegel2010deNovoDieselAlder}
\bibinfo{author}{J.~B. Siegel}, \bibinfo{author}{A.~Zanghellini},
  \bibinfo{author}{H.~M. Lovick}, \bibinfo{author}{G.~Kiss},
  \bibinfo{author}{A.~R. Lambert}, \bibinfo{author}{J.~L. St.~Clair},
  \bibinfo{author}{J.~L. Gallaher}, \bibinfo{author}{D.~Hilvert},
  \bibinfo{author}{M.~H. Gelb}, \bibinfo{author}{B.~L. Stoddard}, et~al.,
\newblock \bibinfo{title}{Computational design of an enzyme catalyst for a
  stereoselective bimolecular diels-alder reaction},
\newblock \bibinfo{journal}{Science} \bibinfo{volume}{329}
  (\bibinfo{year}{2010}) \bibinfo{pages}{309--313}.
%Type = Article
\bibitem[{Richter et~al.(2012)Richter, Blomberg, Khare, Kiss, Kuzin, Smith,
  Gallaher, Pianowski, Helgeson, Grjasnow et~al.}]{richter2012esteraseDesign}
\bibinfo{author}{F.~Richter}, \bibinfo{author}{R.~Blomberg},
  \bibinfo{author}{S.~D. Khare}, \bibinfo{author}{G.~Kiss},
  \bibinfo{author}{A.~P. Kuzin}, \bibinfo{author}{A.~J. Smith},
  \bibinfo{author}{J.~Gallaher}, \bibinfo{author}{Z.~Pianowski},
  \bibinfo{author}{R.~C. Helgeson}, \bibinfo{author}{A.~Grjasnow}, et~al.,
\newblock \bibinfo{title}{Computational design of catalytic dyads and oxyanion
  holes for ester hydrolysis},
\newblock \bibinfo{journal}{Journal of the American Chemical Society}
  \bibinfo{volume}{134} (\bibinfo{year}{2012}) \bibinfo{pages}{16197--16206}.
%Type = Article
\bibitem[{Ruiz-Pern{\'\i}a et~al.(2024)Ruiz-Pern{\'\i}a, Swiderek, Bertran,
  Moliner, and Tu{\~n}{\'o}n}]{ruiz2024electrostatics}
\bibinfo{author}{J.~J. Ruiz-Pern{\'\i}a}, \bibinfo{author}{K.~Swiderek},
  \bibinfo{author}{J.~Bertran}, \bibinfo{author}{V.~Moliner},
  \bibinfo{author}{I.~Tu{\~n}{\'o}n},
\newblock \bibinfo{title}{Electrostatics as a guiding principle in
  understanding and designing enzymes},
\newblock \bibinfo{journal}{Journal of Chemical Theory and Computation}
  \bibinfo{volume}{20} (\bibinfo{year}{2024}) \bibinfo{pages}{1783--1795}.
%Type = Article
\bibitem[{Hammes-Schiffer and Benkovic(2006)}]{hammes2006Dynamcis}
\bibinfo{author}{S.~Hammes-Schiffer}, \bibinfo{author}{S.~J. Benkovic},
\newblock \bibinfo{title}{Relating protein motion to catalysis},
\newblock \bibinfo{journal}{Annu. Rev. Biochem.} \bibinfo{volume}{75}
  (\bibinfo{year}{2006}) \bibinfo{pages}{519--541}.
%Type = Article
\bibitem[{Romero-Romero et~al.(2023)Romero-Romero, Lindner, and
  Ferruz}]{romero2023exploringSequenceSpaceWithGenerativeModels}
\bibinfo{author}{S.~Romero-Romero}, \bibinfo{author}{S.~Lindner},
  \bibinfo{author}{N.~Ferruz},
\newblock \bibinfo{title}{Exploring the protein sequence space with global
  generative models},
\newblock \bibinfo{journal}{Cold Spring Harbor Perspectives in Biology}
  \bibinfo{volume}{15} (\bibinfo{year}{2023}) \bibinfo{pages}{a041471}.
%Type = Article
\bibitem[{Bepler and Berger(2021)}]{bepler2021learningTheProteinLanguage}
\bibinfo{author}{T.~Bepler}, \bibinfo{author}{B.~Berger},
\newblock \bibinfo{title}{Learning the protein language: Evolution, structure,
  and function},
\newblock \bibinfo{journal}{Cell systems} \bibinfo{volume}{12}
  (\bibinfo{year}{2021}) \bibinfo{pages}{654--669}.
%Type = Article
\bibitem[{Albanese et~al.(2025)Albanese, Barbe, Tagami, Woolfson, and
  Schiex}]{albanese2025computationalProteinDesign}
\bibinfo{author}{K.~I. Albanese}, \bibinfo{author}{S.~Barbe},
  \bibinfo{author}{S.~Tagami}, \bibinfo{author}{D.~N. Woolfson},
  \bibinfo{author}{T.~Schiex},
\newblock \bibinfo{title}{Computational protein design},
\newblock \bibinfo{journal}{Nature Reviews Methods Primers} \bibinfo{volume}{5}
  (\bibinfo{year}{2025}) \bibinfo{pages}{13}.
%Type = Article
\bibitem[{Yang et~al.(2024)Yang, Li, and Arnold}]{yang2024MLDE_Review}
\bibinfo{author}{J.~Yang}, \bibinfo{author}{F.-Z. Li}, \bibinfo{author}{F.~H.
  Arnold},
\newblock \bibinfo{title}{Opportunities and challenges for machine
  learning-assisted enzyme engineering},
\newblock \bibinfo{journal}{ACS Central Science} \bibinfo{volume}{10}
  (\bibinfo{year}{2024}) \bibinfo{pages}{226--241}.
%Type = Article
\bibitem[{Maguire et~al.(2024)Maguire, Bloznelyte, Adepoju, Armean-Jones,
  Dewan, Fozzard, Gupta, Ibrahimi, Jones, Lalli
  et~al.}]{maguire2024ESM_cutinase_petase}
\bibinfo{author}{R.~Maguire}, \bibinfo{author}{K.~Bloznelyte},
  \bibinfo{author}{F.~Adepoju}, \bibinfo{author}{M.~Armean-Jones},
  \bibinfo{author}{S.~Dewan}, \bibinfo{author}{S.~Fozzard},
  \bibinfo{author}{A.~Gupta}, \bibinfo{author}{E.~Ibrahimi},
  \bibinfo{author}{F.~P. Jones}, \bibinfo{author}{P.~Lalli}, et~al.,
\newblock \bibinfo{title}{Protein language models in directed evolution},
\newblock \bibinfo{journal}{bioRxiv}  (\bibinfo{year}{2024})
  \bibinfo{pages}{2024--08}.
%Type = Article
\bibitem[{Singh et~al.(2025)Singh, Lane, Yu, Lu, Ramos, Cui, and
  Zhao}]{singh2025ESM_atHMT_phytases}
\bibinfo{author}{N.~Singh}, \bibinfo{author}{S.~Lane}, \bibinfo{author}{T.~Yu},
  \bibinfo{author}{J.~Lu}, \bibinfo{author}{A.~Ramos},
  \bibinfo{author}{H.~Cui}, \bibinfo{author}{H.~Zhao},
\newblock \bibinfo{title}{A generalized platform for artificial
  intelligence-powered autonomous enzyme engineering},
\newblock \bibinfo{journal}{Nature Communications} \bibinfo{volume}{16}
  (\bibinfo{year}{2025}) \bibinfo{pages}{5648}.
  \DOIprefix\doi{10.1038/s41467-025-61209-y}.
%Type = Article
\bibitem[{Lin et~al.(2023)Lin, Akin, Rao, Hie, Zhu, Lu, Smetanin, Verkuil,
  Kabeli, Shmueli et~al.}]{lin2023ESM2}
\bibinfo{author}{Z.~Lin}, \bibinfo{author}{H.~Akin}, \bibinfo{author}{R.~Rao},
  \bibinfo{author}{B.~Hie}, \bibinfo{author}{Z.~Zhu}, \bibinfo{author}{W.~Lu},
  \bibinfo{author}{N.~Smetanin}, \bibinfo{author}{R.~Verkuil},
  \bibinfo{author}{O.~Kabeli}, \bibinfo{author}{Y.~Shmueli}, et~al.,
\newblock \bibinfo{title}{Evolutionary-scale prediction of atomic-level protein
  structure with a language model},
\newblock \bibinfo{journal}{Science} \bibinfo{volume}{379}
  (\bibinfo{year}{2023}) \bibinfo{pages}{1123--1130}.
%Type = Article
\bibitem[{Perrotta et~al.(2024)Perrotta, Vinke, Ferreira, Moret, Mahas,
  Chiappino-Pepe, Riedmayr, Mehra, Lehmann, and
  Church}]{perrotta2024ESM_deaminase_church}
\bibinfo{author}{R.~M. Perrotta}, \bibinfo{author}{S.~Vinke},
  \bibinfo{author}{R.~Ferreira}, \bibinfo{author}{M.~Moret},
  \bibinfo{author}{A.~Mahas}, \bibinfo{author}{A.~Chiappino-Pepe},
  \bibinfo{author}{L.~M. Riedmayr}, \bibinfo{author}{A.-T. Mehra},
  \bibinfo{author}{L.~S. Lehmann}, \bibinfo{author}{G.~M. Church},
\newblock \bibinfo{title}{Machine learning and directed evolution of base
  editing enzymes},
\newblock \bibinfo{journal}{bioRxiv}  (\bibinfo{year}{2024})
  \bibinfo{pages}{2024--05}.
%Type = Article
\bibitem[{He et~al.(2024)He, Zhou, Chang, Chen, Liu, Li, Fan, Sun, Miao, Huang
  et~al.}]{he2024ESM_glycosylase}
\bibinfo{author}{Y.~He}, \bibinfo{author}{X.~Zhou}, \bibinfo{author}{C.~Chang},
  \bibinfo{author}{G.~Chen}, \bibinfo{author}{W.~Liu}, \bibinfo{author}{G.~Li},
  \bibinfo{author}{X.~Fan}, \bibinfo{author}{M.~Sun},
  \bibinfo{author}{C.~Miao}, \bibinfo{author}{Q.~Huang}, et~al.,
\newblock \bibinfo{title}{Protein language models-assisted optimization of a
  uracil-n-glycosylase variant enables programmable t-to-g and t-to-c base
  editing},
\newblock \bibinfo{journal}{Molecular Cell} \bibinfo{volume}{84}
  (\bibinfo{year}{2024}) \bibinfo{pages}{1257--1270}.
%Type = Article
\bibitem[{Wei et~al.(2025)Wei, Cheng, Song, Liu, Xu, Huang, Wang, Zhang, Shu,
  and Wei}]{wei2025ProMEP_Cas9}
\bibinfo{author}{D.~Wei}, \bibinfo{author}{P.~Cheng},
  \bibinfo{author}{Z.~Song}, \bibinfo{author}{Y.~Liu}, \bibinfo{author}{X.~Xu},
  \bibinfo{author}{X.~Huang}, \bibinfo{author}{X.~Wang},
  \bibinfo{author}{Y.~Zhang}, \bibinfo{author}{W.~Shu},
  \bibinfo{author}{Y.~Wei},
\newblock \bibinfo{title}{Ai-guided cas9 engineering provides an effective
  strategy to enhance base editing},
\newblock \bibinfo{journal}{Molecular Systems Biology} \bibinfo{volume}{21}
  (\bibinfo{year}{2025}) \bibinfo{pages}{1563--1580}.
%Type = Article
\bibitem[{Rives et~al.(2021)Rives, Meier, Sercu, Goyal, Lin, Liu, Guo, Ott,
  Zitnick, Ma et~al.}]{rives2021ESM1b}
\bibinfo{author}{A.~Rives}, \bibinfo{author}{J.~Meier},
  \bibinfo{author}{T.~Sercu}, \bibinfo{author}{S.~Goyal},
  \bibinfo{author}{Z.~Lin}, \bibinfo{author}{J.~Liu}, \bibinfo{author}{D.~Guo},
  \bibinfo{author}{M.~Ott}, \bibinfo{author}{C.~L. Zitnick},
  \bibinfo{author}{J.~Ma}, et~al.,
\newblock \bibinfo{title}{Biological structure and function emerge from scaling
  unsupervised learning to 250 million protein sequences},
\newblock \bibinfo{journal}{Proceedings of the National Academy of Sciences}
  \bibinfo{volume}{118} (\bibinfo{year}{2021}) \bibinfo{pages}{e2016239118}.
%Type = Article
\bibitem[{Meier et~al.(2021)Meier, Rao, Verkuil, Liu, Sercu, and
  Rives}]{meier2021ESM1v}
\bibinfo{author}{J.~Meier}, \bibinfo{author}{R.~Rao},
  \bibinfo{author}{R.~Verkuil}, \bibinfo{author}{J.~Liu},
  \bibinfo{author}{T.~Sercu}, \bibinfo{author}{A.~Rives},
\newblock \bibinfo{title}{Language models enable zero-shot prediction of the
  effects of mutations on protein function},
\newblock \bibinfo{journal}{Advances in neural information processing systems}
  \bibinfo{volume}{34} (\bibinfo{year}{2021}) \bibinfo{pages}{29287--29303}.
%Type = Article
\bibitem[{Rao et~al.(2021)Rao, Liu, Verkuil, Meier, Canny, Abbeel, Sercu, and
  Rives}]{rao2021MSATransformer}
\bibinfo{author}{R.~M. Rao}, \bibinfo{author}{J.~Liu},
  \bibinfo{author}{R.~Verkuil}, \bibinfo{author}{J.~Meier},
  \bibinfo{author}{J.~Canny}, \bibinfo{author}{P.~Abbeel},
  \bibinfo{author}{T.~Sercu}, \bibinfo{author}{A.~Rives},
\newblock \bibinfo{title}{Msa transformer},
\newblock \bibinfo{journal}{International conference on machine learning}
  (\bibinfo{year}{2021}) \bibinfo{pages}{8844--8856}.
%Type = Article
\bibitem[{Notin et~al.(2023)Notin, Kollasch, Ritter, Van~Niekerk, Paul,
  Spinner, Rollins, Shaw, Orenbuch, Weitzman et~al.}]{notin2023proteingym}
\bibinfo{author}{P.~Notin}, \bibinfo{author}{A.~Kollasch},
  \bibinfo{author}{D.~Ritter}, \bibinfo{author}{L.~Van~Niekerk},
  \bibinfo{author}{S.~Paul}, \bibinfo{author}{H.~Spinner},
  \bibinfo{author}{N.~Rollins}, \bibinfo{author}{A.~Shaw},
  \bibinfo{author}{R.~Orenbuch}, \bibinfo{author}{R.~Weitzman}, et~al.,
\newblock \bibinfo{title}{Proteingym: Large-scale benchmarks for protein
  fitness prediction and design},
\newblock \bibinfo{journal}{Advances in Neural Information Processing Systems}
  \bibinfo{volume}{36} (\bibinfo{year}{2023}) \bibinfo{pages}{64331--64379}.
%Type = Article
\bibitem[{Ferruz et~al.(2022)Ferruz, Schmidt, and
  H{\"o}cker}]{ferruz2022protgpt2}
\bibinfo{author}{N.~Ferruz}, \bibinfo{author}{S.~Schmidt},
  \bibinfo{author}{B.~H{\"o}cker},
\newblock \bibinfo{title}{Protgpt2 is a deep unsupervised language model for
  protein design},
\newblock \bibinfo{journal}{Nature communications} \bibinfo{volume}{13}
  (\bibinfo{year}{2022}) \bibinfo{pages}{4348}.
%Type = Article
\bibitem[{Madani et~al.(2023)Madani, Krause, Greene, Subramanian, Mohr, Holton,
  Olmos~Jr, Xiong, Sun, Socher et~al.}]{madani2023Progen}
\bibinfo{author}{A.~Madani}, \bibinfo{author}{B.~Krause},
  \bibinfo{author}{E.~R. Greene}, \bibinfo{author}{S.~Subramanian},
  \bibinfo{author}{B.~P. Mohr}, \bibinfo{author}{J.~M. Holton},
  \bibinfo{author}{J.~L. Olmos~Jr}, \bibinfo{author}{C.~Xiong},
  \bibinfo{author}{Z.~Z. Sun}, \bibinfo{author}{R.~Socher}, et~al.,
\newblock \bibinfo{title}{Large language models generate functional protein
  sequences across diverse families},
\newblock \bibinfo{journal}{Nature biotechnology} \bibinfo{volume}{41}
  (\bibinfo{year}{2023}) \bibinfo{pages}{1099--1106}.
%Type = Article
\bibitem[{Nijkamp et~al.(2023)Nijkamp, Ruffolo, Weinstein, Naik, and
  Madani}]{nijkamp2023progen2}
\bibinfo{author}{E.~Nijkamp}, \bibinfo{author}{J.~A. Ruffolo},
  \bibinfo{author}{E.~N. Weinstein}, \bibinfo{author}{N.~Naik},
  \bibinfo{author}{A.~Madani},
\newblock \bibinfo{title}{Progen2: exploring the boundaries of protein language
  models},
\newblock \bibinfo{journal}{Cell systems} \bibinfo{volume}{14}
  (\bibinfo{year}{2023}) \bibinfo{pages}{968--978}.
%Type = Article
\bibitem[{Munsamy et~al.(2024)Munsamy, Illanes-Vicioso, Funcillo, Nakou,
  Lindner, Ayres, Sheehan, Moss, Eckhard, Lorenz et~al.}]{munsamy2024ZymCTRL}
\bibinfo{author}{G.~Munsamy}, \bibinfo{author}{R.~Illanes-Vicioso},
  \bibinfo{author}{S.~Funcillo}, \bibinfo{author}{I.~T. Nakou},
  \bibinfo{author}{S.~Lindner}, \bibinfo{author}{G.~Ayres},
  \bibinfo{author}{L.~S. Sheehan}, \bibinfo{author}{S.~Moss},
  \bibinfo{author}{U.~Eckhard}, \bibinfo{author}{P.~Lorenz}, et~al.,
\newblock \bibinfo{title}{Conditional language models enable the efficient
  design of proficient enzymes},
\newblock \bibinfo{journal}{bioRxiv}  (\bibinfo{year}{2024})
  \bibinfo{pages}{2024--05}.
%Type = Article
\bibitem[{Zvyagin et~al.(2023)Zvyagin, Brace, Hippe, Deng, Zhang, Bohorquez,
  Clyde, Kale, Perez-Rivera, Ma et~al.}]{zvyagin2023GenSLM}
\bibinfo{author}{M.~Zvyagin}, \bibinfo{author}{A.~Brace},
  \bibinfo{author}{K.~Hippe}, \bibinfo{author}{Y.~Deng},
  \bibinfo{author}{B.~Zhang}, \bibinfo{author}{C.~O. Bohorquez},
  \bibinfo{author}{A.~Clyde}, \bibinfo{author}{B.~Kale},
  \bibinfo{author}{D.~Perez-Rivera}, \bibinfo{author}{H.~Ma}, et~al.,
\newblock \bibinfo{title}{Genslms: Genome-scale language models reveal
  sars-cov-2 evolutionary dynamics},
\newblock \bibinfo{journal}{The International Journal of High Performance
  Computing Applications} \bibinfo{volume}{37} (\bibinfo{year}{2023})
  \bibinfo{pages}{683--705}.
%Type = Article
\bibitem[{Nguyen et~al.(2024)Nguyen, Poli, Durrant, Kang, Katrekar, Li, Bartie,
  Thomas, King, Brixi et~al.}]{nguyen2024Evo}
\bibinfo{author}{E.~Nguyen}, \bibinfo{author}{M.~Poli}, \bibinfo{author}{M.~G.
  Durrant}, \bibinfo{author}{B.~Kang}, \bibinfo{author}{D.~Katrekar},
  \bibinfo{author}{D.~B. Li}, \bibinfo{author}{L.~J. Bartie},
  \bibinfo{author}{A.~W. Thomas}, \bibinfo{author}{S.~H. King},
  \bibinfo{author}{G.~Brixi}, et~al.,
\newblock \bibinfo{title}{Sequence modeling and design from molecular to genome
  scale with evo},
\newblock \bibinfo{journal}{Science} \bibinfo{volume}{386}
  (\bibinfo{year}{2024}) \bibinfo{pages}{eado9336}.
%Type = Article
\bibitem[{Dai et~al.(2024)Dai, You, Zhu, Gao, Fu, Zhou, Su, Wang, Fan, Ma
  et~al.}]{dai2024Pinal}
\bibinfo{author}{F.~Dai}, \bibinfo{author}{S.~You}, \bibinfo{author}{Y.~Zhu},
  \bibinfo{author}{Y.~Gao}, \bibinfo{author}{L.~Fu}, \bibinfo{author}{X.~Zhou},
  \bibinfo{author}{J.~Su}, \bibinfo{author}{C.~Wang}, \bibinfo{author}{Y.~Fan},
  \bibinfo{author}{X.~Ma}, et~al.,
\newblock \bibinfo{title}{Toward de novo protein design from natural language},
\newblock \bibinfo{journal}{BioRxiv}  (\bibinfo{year}{2024})
  \bibinfo{pages}{2024--08}.
%Type = Article
\bibitem[{Romero-Romero et~al.(2024)Romero-Romero, Braun, Kossendey, Ferruz,
  Schmidt, and H{\"o}cker}]{romero2024ProtGPT_ZymCTRL_TIM}
\bibinfo{author}{S.~Romero-Romero}, \bibinfo{author}{A.~E. Braun},
  \bibinfo{author}{T.~Kossendey}, \bibinfo{author}{N.~Ferruz},
  \bibinfo{author}{S.~Schmidt}, \bibinfo{author}{B.~H{\"o}cker},
\newblock \bibinfo{title}{De novo design of triosephosphate isomerases using
  generative language models},
\newblock \bibinfo{journal}{bioRxiv}  (\bibinfo{year}{2024})
  \bibinfo{pages}{2024--11}.
%Type = Article
\bibitem[{Armer et~al.(2025)Armer, Kane, Cortade, Redestig, Estell, Yusuf,
  Rollins, Spinner, Marks, Brunette
  et~al.}]{armer2025AlignFoundation_AlphaAmylases}
\bibinfo{author}{C.~Armer}, \bibinfo{author}{H.~Kane}, \bibinfo{author}{D.~L.
  Cortade}, \bibinfo{author}{H.~Redestig}, \bibinfo{author}{D.~A. Estell},
  \bibinfo{author}{A.~Yusuf}, \bibinfo{author}{N.~Rollins},
  \bibinfo{author}{A.~Spinner}, \bibinfo{author}{D.~Marks},
  \bibinfo{author}{T.~Brunette}, et~al.,
\newblock \bibinfo{title}{Results of the protein engineering tournament: an
  open science benchmark for protein modeling and design},
\newblock \bibinfo{journal}{Proteins: Structure, Function, and Bioinformatics}
  \bibinfo{volume}{93} (\bibinfo{year}{2025}) \bibinfo{pages}{2005--2014}.
%Type = Article
\bibitem[{Ivan{\v{c}}i{\'c} et~al.(2025)Ivan{\v{c}}i{\'c}, Agudelo,
  Lindstrom-Vautrin, Jaraba-Wallace, Gallo, Das, Ragel, Herrero-Vicente,
  Higueras, Billeci et~al.}]{ivanvcic2025Progen_Piggybac}
\bibinfo{author}{D.~Ivan{\v{c}}i{\'c}}, \bibinfo{author}{A.~Agudelo},
  \bibinfo{author}{J.~Lindstrom-Vautrin}, \bibinfo{author}{J.~Jaraba-Wallace},
  \bibinfo{author}{M.~Gallo}, \bibinfo{author}{R.~Das},
  \bibinfo{author}{A.~Ragel}, \bibinfo{author}{J.~Herrero-Vicente},
  \bibinfo{author}{I.~Higueras}, \bibinfo{author}{F.~Billeci}, et~al.,
\newblock \bibinfo{title}{Discovery and protein language model-guided design of
  hyperactive transposases},
\newblock \bibinfo{journal}{Nature Biotechnology}  (\bibinfo{year}{2025})
  \bibinfo{pages}{1--6}.
%Type = Article
\bibitem[{Stocco et~al.(2024)Stocco, Artigues-Lleixa, Hunklinger, Widatalla,
  Guell, and Ferruz}]{stocco2024ProtRL}
\bibinfo{author}{F.~Stocco}, \bibinfo{author}{M.~Artigues-Lleixa},
  \bibinfo{author}{A.~Hunklinger}, \bibinfo{author}{T.~Widatalla},
  \bibinfo{author}{M.~Guell}, \bibinfo{author}{N.~Ferruz},
\newblock \bibinfo{title}{Guiding generative protein language models with
  reinforcement learning},
\newblock \bibinfo{journal}{arXiv preprint arXiv:2412.12979}
  (\bibinfo{year}{2024}).
%Type = Article
\bibitem[{Stocco et~al.(2025)Stocco, Garibbo, and
  Ferruz}]{stocco2025Steering_review}
\bibinfo{author}{F.~Stocco}, \bibinfo{author}{M.~Garibbo},
  \bibinfo{author}{N.~Ferruz},
\newblock \bibinfo{title}{Guiding generative models for protein design:
  Prompting, steering and aligning},
\newblock \bibinfo{journal}{arXiv preprint arXiv:2511.21476}
  (\bibinfo{year}{2025}).
%Type = Article
\bibitem[{Dauparas et~al.(2022)Dauparas, Anishchenko, Bennett, Bai, Ragotte,
  Milles, Wicky, Courbet, de~Haas, Bethel et~al.}]{dauparas2022ProteinMPNN}
\bibinfo{author}{J.~Dauparas}, \bibinfo{author}{I.~Anishchenko},
  \bibinfo{author}{N.~Bennett}, \bibinfo{author}{H.~Bai},
  \bibinfo{author}{R.~J. Ragotte}, \bibinfo{author}{L.~F. Milles},
  \bibinfo{author}{B.~I. Wicky}, \bibinfo{author}{A.~Courbet},
  \bibinfo{author}{R.~J. de~Haas}, \bibinfo{author}{N.~Bethel}, et~al.,
\newblock \bibinfo{title}{Robust deep learning--based protein sequence design
  using proteinmpnn},
\newblock \bibinfo{journal}{Science} \bibinfo{volume}{378}
  (\bibinfo{year}{2022}) \bibinfo{pages}{49--56}.
%Type = Article
\bibitem[{Dauparas et~al.(2025)Dauparas, Lee, Pecoraro, An, Anishchenko,
  Glasscock, and Baker}]{dauparas2025LigandMPNN}
\bibinfo{author}{J.~Dauparas}, \bibinfo{author}{G.~R. Lee},
  \bibinfo{author}{R.~Pecoraro}, \bibinfo{author}{L.~An},
  \bibinfo{author}{I.~Anishchenko}, \bibinfo{author}{C.~Glasscock},
  \bibinfo{author}{D.~Baker},
\newblock \bibinfo{title}{Atomic context-conditioned protein sequence design
  using ligandmpnn},
\newblock \bibinfo{journal}{Nature Methods}  (\bibinfo{year}{2025})
  \bibinfo{pages}{1--7}.
%Type = Article
\bibitem[{Liu et~al.(2025)Liu, You, Guo, Xu, Zhang, and
  Lai}]{liu2025geoevobuilder}
\bibinfo{author}{J.~Liu}, \bibinfo{author}{H.~You}, \bibinfo{author}{Z.~Guo},
  \bibinfo{author}{Q.~Xu}, \bibinfo{author}{C.~Zhang},
  \bibinfo{author}{L.~Lai},
\newblock \bibinfo{title}{Geoevobuilder: A deep learning framework for
  efficient functional and thermostable protein design},
\newblock \bibinfo{journal}{Proceedings of the National Academy of Sciences}
  \bibinfo{volume}{122} (\bibinfo{year}{2025}) \bibinfo{pages}{e2504117122}.
%Type = Article
\bibitem[{Sumida et~al.(2024)Sumida, N{\'u}{\~n}ez-Franco, Kalvet, Pellock,
  Wicky, Milles, Dauparas, Wang, Kipnis, Jameson
  et~al.}]{sumida2024ProteinMPNN_TEV}
\bibinfo{author}{K.~H. Sumida}, \bibinfo{author}{R.~N{\'u}{\~n}ez-Franco},
  \bibinfo{author}{I.~Kalvet}, \bibinfo{author}{S.~J. Pellock},
  \bibinfo{author}{B.~I. Wicky}, \bibinfo{author}{L.~F. Milles},
  \bibinfo{author}{J.~Dauparas}, \bibinfo{author}{J.~Wang},
  \bibinfo{author}{Y.~Kipnis}, \bibinfo{author}{N.~Jameson}, et~al.,
\newblock \bibinfo{title}{Improving protein expression, stability, and function
  with proteinmpnn},
\newblock \bibinfo{journal}{Journal of the American Chemical Society}
  \bibinfo{volume}{146} (\bibinfo{year}{2024}) \bibinfo{pages}{2054--2061}.
%Type = Article
\bibitem[{King et~al.(2025)King, Sumida, Caruso, Baker, and
  Zalatan}]{king2025ProteinMPNN_aKG}
\bibinfo{author}{B.~R. King}, \bibinfo{author}{K.~H. Sumida},
  \bibinfo{author}{J.~L. Caruso}, \bibinfo{author}{D.~Baker},
  \bibinfo{author}{J.~G. Zalatan},
\newblock \bibinfo{title}{Computational stabilization of a non-heme iron enzyme
  enables efficient evolution of new function},
\newblock \bibinfo{journal}{Angewandte Chemie International Edition}
  \bibinfo{volume}{64} (\bibinfo{year}{2025}) \bibinfo{pages}{e202414705}.
%Type = Article
\bibitem[{Bloom et~al.(2006)Bloom, Labthavikul, Otey, and
  Arnold}]{bloom2006protein_stabillity_promotes_evolvability}
\bibinfo{author}{J.~D. Bloom}, \bibinfo{author}{S.~T. Labthavikul},
  \bibinfo{author}{C.~R. Otey}, \bibinfo{author}{F.~H. Arnold},
\newblock \bibinfo{title}{Protein stability promotes evolvability},
\newblock \bibinfo{journal}{Proceedings of the National Academy of Sciences}
  \bibinfo{volume}{103} (\bibinfo{year}{2006}) \bibinfo{pages}{5869--5874}.
%Type = Article
\bibitem[{Tokuriki et~al.(2008)Tokuriki, Stricher, Serrano, and
  Tawfik}]{tokuriki2008stabillity_function_tradeoff}
\bibinfo{author}{N.~Tokuriki}, \bibinfo{author}{F.~Stricher},
  \bibinfo{author}{L.~Serrano}, \bibinfo{author}{D.~S. Tawfik},
\newblock \bibinfo{title}{How protein stability and new functions trade off},
\newblock \bibinfo{journal}{PLoS computational biology} \bibinfo{volume}{4}
  (\bibinfo{year}{2008}) \bibinfo{pages}{e1000002}.
%Type = Article
\bibitem[{King et~al.(2025)King, Sumida, Caruso, Baker, and
  Zalatan}]{king2025ProteinMPNN_tP4H}
\bibinfo{author}{B.~R. King}, \bibinfo{author}{K.~H. Sumida},
  \bibinfo{author}{J.~L. Caruso}, \bibinfo{author}{D.~Baker},
  \bibinfo{author}{J.~G. Zalatan},
\newblock \bibinfo{title}{Computational stabilization of a non-heme iron enzyme
  enables efficient evolution of new function},
\newblock \bibinfo{journal}{Angewandte Chemie International Edition}
  \bibinfo{volume}{64} (\bibinfo{year}{2025}) \bibinfo{pages}{e202414705}.
%Type = Article
\bibitem[{Merlicek et~al.(2025)Merlicek, Neumann, Lear, Degiorgi, de~Waal,
  Cotet, Mulholland, and Bunzel}]{merlicek2025aizymes}
\bibinfo{author}{L.~P. Merlicek}, \bibinfo{author}{J.~Neumann},
  \bibinfo{author}{A.~Lear}, \bibinfo{author}{V.~Degiorgi},
  \bibinfo{author}{M.~M. de~Waal}, \bibinfo{author}{T.-S. Cotet},
  \bibinfo{author}{A.~J. Mulholland}, \bibinfo{author}{H.~A. Bunzel},
\newblock \bibinfo{title}{Ai. zymes: A modular platform for evolutionary enzyme
  design},
\newblock \bibinfo{journal}{Angewandte Chemie International Edition}
  (\bibinfo{year}{2025}) \bibinfo{pages}{e202507031}.
%Type = Article
\bibitem[{Sun et~al.(2025)Sun, Li, Ji, Schwaneberg, and Li}]{sun2025cmdmpnn}
\bibinfo{author}{C.-q. Sun}, \bibinfo{author}{Z.-m. Li},
  \bibinfo{author}{Y.~Ji}, \bibinfo{author}{U.~Schwaneberg},
  \bibinfo{author}{Z.-l. Li},
\newblock \bibinfo{title}{Cmdmpnn: Combining comparative molecular dynamics and
  proteinmpnn to rapidly expand enzyme substrate spectrum},
\newblock \bibinfo{journal}{Journal of Chemical Information and Modeling}
  \bibinfo{volume}{65} (\bibinfo{year}{2025}) \bibinfo{pages}{2741--2747}.
%Type = Article
\bibitem[{Hou et~al.(2025)Hou, Huang, Qi, Tugwell, Alturaifi, Chen, Zhang, Lu,
  Mann, Liu et~al.}]{hou2025denovo_porphyrin_containing_proteins}
\bibinfo{author}{K.~Hou}, \bibinfo{author}{W.~Huang}, \bibinfo{author}{M.~Qi},
  \bibinfo{author}{T.~H. Tugwell}, \bibinfo{author}{T.~M. Alturaifi},
  \bibinfo{author}{Y.~Chen}, \bibinfo{author}{X.~Zhang},
  \bibinfo{author}{L.~Lu}, \bibinfo{author}{S.~I. Mann},
  \bibinfo{author}{P.~Liu}, et~al.,
\newblock \bibinfo{title}{De novo design of porphyrin-containing proteins as
  efficient and stereoselective catalysts},
\newblock \bibinfo{journal}{Science} \bibinfo{volume}{388}
  (\bibinfo{year}{2025}) \bibinfo{pages}{665--670}.
%Type = Article
\bibitem[{Huang et~al.(2025)Huang, Adornato, Horst, Alturaifi, Hou, Liu,
  DeGrado, and Yang}]{huang2025denovo_GeH_insertion}
\bibinfo{author}{W.~Huang}, \bibinfo{author}{G.~M. Adornato},
  \bibinfo{author}{M.~Horst}, \bibinfo{author}{T.~M. Alturaifi},
  \bibinfo{author}{K.~Hou}, \bibinfo{author}{P.~Liu}, \bibinfo{author}{W.~F.
  DeGrado}, \bibinfo{author}{Y.~Yang},
\newblock \bibinfo{title}{De novo design, directed evolution and computational
  study of heme-binding helical bundle protein catalysts for biocatalytic
  enantioselective ge--h insertion},
\newblock \bibinfo{journal}{Journal of the American Chemical Society}
  \bibinfo{volume}{147} (\bibinfo{year}{2025}) \bibinfo{pages}{40869--40878}.
%Type = Article
\bibitem[{Watson et~al.(2023)Watson, Juergens, Bennett, Trippe, Yim, Eisenach,
  Ahern, Borst, Ragotte, Milles et~al.}]{watson2023RFdiffusion}
\bibinfo{author}{J.~L. Watson}, \bibinfo{author}{D.~Juergens},
  \bibinfo{author}{N.~R. Bennett}, \bibinfo{author}{B.~L. Trippe},
  \bibinfo{author}{J.~Yim}, \bibinfo{author}{H.~E. Eisenach},
  \bibinfo{author}{W.~Ahern}, \bibinfo{author}{A.~J. Borst},
  \bibinfo{author}{R.~J. Ragotte}, \bibinfo{author}{L.~F. Milles}, et~al.,
\newblock \bibinfo{title}{De novo design of protein structure and function with
  rfdiffusion},
\newblock \bibinfo{journal}{Nature} \bibinfo{volume}{620}
  (\bibinfo{year}{2023}) \bibinfo{pages}{1089--1100}.
%Type = Article
\bibitem[{Ahern et~al.(2025)Ahern, Yim, Tischer, Salike, Woodbury, Kim, Kalvet,
  Kipnis, Coventry, Altae-Tran et~al.}]{ahern2025RFDiffusion2}
\bibinfo{author}{W.~Ahern}, \bibinfo{author}{J.~Yim},
  \bibinfo{author}{D.~Tischer}, \bibinfo{author}{S.~Salike},
  \bibinfo{author}{S.~M. Woodbury}, \bibinfo{author}{D.~Kim},
  \bibinfo{author}{I.~Kalvet}, \bibinfo{author}{Y.~Kipnis},
  \bibinfo{author}{B.~Coventry}, \bibinfo{author}{H.~R. Altae-Tran}, et~al.,
\newblock \bibinfo{title}{Atom-level enzyme active site scaffolding using
  rfdiffusion2},
\newblock \bibinfo{journal}{Nature Methods}  (\bibinfo{year}{2025})
  \bibinfo{pages}{1--10}.
%Type = Article
\bibitem[{Butcher et~al.(2025)Butcher, Krishna, Mitra, Brent, Li, Corley, Kim,
  Funk, Mathis, Salike et~al.}]{butcher2025RFdiffusion3}
\bibinfo{author}{J.~Butcher}, \bibinfo{author}{R.~Krishna},
  \bibinfo{author}{R.~Mitra}, \bibinfo{author}{R.~I. Brent},
  \bibinfo{author}{Y.~Li}, \bibinfo{author}{N.~Corley}, \bibinfo{author}{P.~T.
  Kim}, \bibinfo{author}{J.~Funk}, \bibinfo{author}{S.~Mathis},
  \bibinfo{author}{S.~Salike}, et~al.,
\newblock \bibinfo{title}{De novo design of all-atom biomolecular interactions
  with rfdiffusion3},
\newblock \bibinfo{journal}{bioRxiv}  (\bibinfo{year}{2025}).
%Type = Article
\bibitem[{Lauko et~al.(2025)Lauko, Pellock, Sumida, Anishchenko, Juergens,
  Ahern, Jeung, Shida, Hunt, Kalvet et~al.}]{lauko2025serin_hydrolase}
\bibinfo{author}{A.~Lauko}, \bibinfo{author}{S.~J. Pellock},
  \bibinfo{author}{K.~H. Sumida}, \bibinfo{author}{I.~Anishchenko},
  \bibinfo{author}{D.~Juergens}, \bibinfo{author}{W.~Ahern},
  \bibinfo{author}{J.~Jeung}, \bibinfo{author}{A.~F. Shida},
  \bibinfo{author}{A.~Hunt}, \bibinfo{author}{I.~Kalvet}, et~al.,
\newblock \bibinfo{title}{Computational design of serine hydrolases},
\newblock \bibinfo{journal}{Science} \bibinfo{volume}{388}
  (\bibinfo{year}{2025}) \bibinfo{pages}{eadu2454}.
%Type = Article
\bibitem[{Bar-Even et~al.(2011)Bar-Even, Noor, Savir, Liebermeister, Davidi,
  Tawfik, and Milo}]{bar2011moderately}
\bibinfo{author}{A.~Bar-Even}, \bibinfo{author}{E.~Noor},
  \bibinfo{author}{Y.~Savir}, \bibinfo{author}{W.~Liebermeister},
  \bibinfo{author}{D.~Davidi}, \bibinfo{author}{D.~S. Tawfik},
  \bibinfo{author}{R.~Milo},
\newblock \bibinfo{title}{The moderately efficient enzyme: evolutionary and
  physicochemical trends shaping enzyme parameters},
\newblock \bibinfo{journal}{Biochemistry} \bibinfo{volume}{50}
  (\bibinfo{year}{2011}) \bibinfo{pages}{4402--4410}.
%Type = Article
\bibitem[{Braun et~al.(2025)Braun, Tripp, Chakatok, Kaltenbrunner, Fischer,
  Stoll, Bijelic, Elaily, Totaro, Moser et~al.}]{braun2025RiffDiff}
\bibinfo{author}{M.~Braun}, \bibinfo{author}{A.~Tripp},
  \bibinfo{author}{M.~Chakatok}, \bibinfo{author}{S.~Kaltenbrunner},
  \bibinfo{author}{C.~Fischer}, \bibinfo{author}{D.~Stoll},
  \bibinfo{author}{A.~Bijelic}, \bibinfo{author}{W.~Elaily},
  \bibinfo{author}{M.~G. Totaro}, \bibinfo{author}{M.~Moser}, et~al.,
\newblock \bibinfo{title}{Computational enzyme design by catalytic motif
  scaffolding},
\newblock \bibinfo{journal}{Nature}  (\bibinfo{year}{2025})
  \bibinfo{pages}{1--9}.
%Type = Article
\bibitem[{Choi et~al.(2025)Choi, Coventry, Bauer, Venkatesh, Chen, Kim, Bera,
  Kang, Nguyen, Joyce et~al.}]{choi2025RFD2_cystein_protease}
\bibinfo{author}{H.~Choi}, \bibinfo{author}{B.~Coventry},
  \bibinfo{author}{M.~Bauer}, \bibinfo{author}{P.~Venkatesh},
  \bibinfo{author}{A.~Chen}, \bibinfo{author}{D.~Kim}, \bibinfo{author}{A.~K.
  Bera}, \bibinfo{author}{A.~Kang}, \bibinfo{author}{H.~Nguyen},
  \bibinfo{author}{E.~Joyce}, et~al.,
\newblock \bibinfo{title}{Computational design of cysteine proteases},
\newblock \bibinfo{journal}{bioRxiv}  (\bibinfo{year}{2025})
  \bibinfo{pages}{2025--11}.
%Type = Article
\bibitem[{Kim et~al.(2025)Kim, Woodbury, Ahern, Tischer, Kang, Joyce, Bera,
  Hanikel, Salike, Krishna et~al.}]{kim2025RFd2_metallohydrolases}
\bibinfo{author}{D.~Kim}, \bibinfo{author}{S.~M. Woodbury},
  \bibinfo{author}{W.~Ahern}, \bibinfo{author}{D.~Tischer},
  \bibinfo{author}{A.~Kang}, \bibinfo{author}{E.~Joyce}, \bibinfo{author}{A.~K.
  Bera}, \bibinfo{author}{N.~Hanikel}, \bibinfo{author}{S.~Salike},
  \bibinfo{author}{R.~Krishna}, et~al.,
\newblock \bibinfo{title}{Computational design of metallohydrolases},
\newblock \bibinfo{journal}{Nature}  (\bibinfo{year}{2025})
  \bibinfo{pages}{1--8}.
%Type = Article
\bibitem[{Qu et~al.(2026)Qu, Wang, Zhu, Wang, and
  Cao}]{qu2026proteus2_metalloprotease}
\bibinfo{author}{Y.~Qu}, \bibinfo{author}{C.~Wang}, \bibinfo{author}{H.~Zhu},
  \bibinfo{author}{Y.~Wang}, \bibinfo{author}{L.~Cao},
\newblock \bibinfo{title}{De novo design of metalloproteases for targeted
  amyloid-$\beta$ cleavage},
\newblock \bibinfo{journal}{bioRxiv}  (\bibinfo{year}{2026})
  \bibinfo{pages}{2026--01}.
%Type = Article
\bibitem[{Hou et~al.(2025)Hou, Liu, Zafar, and
  Shen}]{hou2025understanding_plm_scaling_fitness_prediction}
\bibinfo{author}{C.~Hou}, \bibinfo{author}{D.~Liu}, \bibinfo{author}{A.~Zafar},
  \bibinfo{author}{Y.~Shen},
\newblock \bibinfo{title}{Understanding protein language model scaling on
  mutation effect prediction},
\newblock \bibinfo{journal}{bioRxiv}  (\bibinfo{year}{2025})
  \bibinfo{pages}{2025--04}.
%Type = Article
\bibitem[{Widatalla et~al.(2024)Widatalla, Rafailov, and
  Hie}]{widatalla2024ProteinDPO}
\bibinfo{author}{T.~Widatalla}, \bibinfo{author}{R.~Rafailov},
  \bibinfo{author}{B.~Hie},
\newblock \bibinfo{title}{Aligning protein generative models with experimental
  fitness via direct preference optimization},
\newblock \bibinfo{journal}{bioRxiv}  (\bibinfo{year}{2024})
  \bibinfo{pages}{2024--05}.
%Type = Article
\bibitem[{Blalock et~al.(2025)Blalock, Seshadri, Babbar, Fahlberg, Kulkarni,
  and Romero}]{blalock2025RLXF}
\bibinfo{author}{N.~Blalock}, \bibinfo{author}{S.~Seshadri},
  \bibinfo{author}{A.~Babbar}, \bibinfo{author}{S.~A. Fahlberg},
  \bibinfo{author}{A.~Kulkarni}, \bibinfo{author}{P.~A. Romero},
\newblock \bibinfo{title}{Functional alignment of protein language models via
  reinforcement learning},
\newblock \bibinfo{journal}{bioRxiv}  (\bibinfo{year}{2025})
  \bibinfo{pages}{2025--05}.
%Type = Article
\bibitem[{Bhatnagar et~al.(2025)Bhatnagar, Jain, Beazer, Curran, Hoffnagle,
  Ching, Martyn, Nayfach, Ruffolo, and Madani}]{bhatnagar2025Progen3}
\bibinfo{author}{A.~Bhatnagar}, \bibinfo{author}{S.~Jain},
  \bibinfo{author}{J.~Beazer}, \bibinfo{author}{S.~C. Curran},
  \bibinfo{author}{A.~M. Hoffnagle}, \bibinfo{author}{K.~S. Ching},
  \bibinfo{author}{M.~Martyn}, \bibinfo{author}{S.~Nayfach},
  \bibinfo{author}{J.~A. Ruffolo}, \bibinfo{author}{A.~Madani},
\newblock \bibinfo{title}{Scaling unlocks broader generation and deeper
  functional understanding of proteins},
\newblock \bibinfo{journal}{bioRxiv}  (\bibinfo{year}{2025})
  \bibinfo{pages}{2025--04}.
%Type = Article
\bibitem[{Brixi et~al.(2025)Brixi, Durrant, Ku, Poli, Brockman, Chang,
  Gonzalez, King, Li, Merchant et~al.}]{brixi2025evo2}
\bibinfo{author}{G.~Brixi}, \bibinfo{author}{M.~G. Durrant},
  \bibinfo{author}{J.~Ku}, \bibinfo{author}{M.~Poli},
  \bibinfo{author}{G.~Brockman}, \bibinfo{author}{D.~Chang},
  \bibinfo{author}{G.~A. Gonzalez}, \bibinfo{author}{S.~H. King},
  \bibinfo{author}{D.~B. Li}, \bibinfo{author}{A.~T. Merchant}, et~al.,
\newblock \bibinfo{title}{Genome modeling and design across all domains of life
  with evo 2},
\newblock \bibinfo{journal}{BioRxiv}  (\bibinfo{year}{2025})
  \bibinfo{pages}{2025--02}.
%Type = Article
\bibitem[{Song et~al.(2025)Song, Li, Cui, Zhou, Qiao, Wei, and
  Li}]{song2025ESM_ASR_Petase}
\bibinfo{author}{Y.~Song}, \bibinfo{author}{A.~Li}, \bibinfo{author}{H.~Cui},
  \bibinfo{author}{B.~Zhou}, \bibinfo{author}{J.~Qiao},
  \bibinfo{author}{J.~Wei}, \bibinfo{author}{X.~Li},
\newblock \bibinfo{title}{Protein language model empowered the robust
  {{ASR-driven PET}} hydrolase featured with two {{PET}} binding motifs},
\newblock \bibinfo{journal}{Green Carbon}  (\bibinfo{year}{2025})
  \bibinfo{pages}{S2950155525000321}.
  \DOIprefix\doi{10.1016/j.greenca.2025.03.005}.
%Type = Article
\bibitem[{Egea et~al.(2025)Egea, Delhommel, Mustafa, Leiss-Maier, Klimper,
  Badmann, Heider, Wille, Groll, Sattler et~al.}]{egea2025AI_lanthanide_lyase}
\bibinfo{author}{P.~W. Egea}, \bibinfo{author}{F.~Delhommel},
  \bibinfo{author}{G.~Mustafa}, \bibinfo{author}{F.~Leiss-Maier},
  \bibinfo{author}{L.~Klimper}, \bibinfo{author}{T.~Badmann},
  \bibinfo{author}{A.~Heider}, \bibinfo{author}{I.~Wille},
  \bibinfo{author}{M.~Groll}, \bibinfo{author}{M.~Sattler}, et~al.,
\newblock \bibinfo{title}{Inter-domain flexibility and ai-guided sequence
  optimization enhance de novo enzyme function}  (\bibinfo{year}{2025}).
%Type = Article
\bibitem[{Hu et~al.(2024)Hu, Yu, and
  Ng}]{hu2024grace_denovo_Carbonic_anhydrase}
\bibinfo{author}{R.-E. Hu}, \bibinfo{author}{C.-H. Yu}, \bibinfo{author}{I.-S.
  Ng},
\newblock \bibinfo{title}{Grace: Generative redesign in artificial
  computational enzymology},
\newblock \bibinfo{journal}{ACS Synthetic Biology} \bibinfo{volume}{13}
  (\bibinfo{year}{2024}) \bibinfo{pages}{4154--4164}.
%Type = Article
\bibitem[{Ruffolo et~al.(2025)Ruffolo, Nayfach, Gallagher, Bhatnagar, Beazer,
  Hussain, Russ, Yip, Hill, Pacesa et~al.}]{ruffolo2025OpenCRISPR}
\bibinfo{author}{J.~A. Ruffolo}, \bibinfo{author}{S.~Nayfach},
  \bibinfo{author}{J.~Gallagher}, \bibinfo{author}{A.~Bhatnagar},
  \bibinfo{author}{J.~Beazer}, \bibinfo{author}{R.~Hussain},
  \bibinfo{author}{J.~Russ}, \bibinfo{author}{J.~Yip},
  \bibinfo{author}{E.~Hill}, \bibinfo{author}{M.~Pacesa}, et~al.,
\newblock \bibinfo{title}{Design of highly functional genome editors by
  modelling crispr--cas sequences},
\newblock \bibinfo{journal}{Nature} \bibinfo{volume}{645}
  (\bibinfo{year}{2025}) \bibinfo{pages}{518--525}.
%Type = Article
\bibitem[{Xu et~al.(2025)Xu, Chen, Meng, Pan, Yan, Li, and
  Li}]{xu2025ProteinMPNN_catalase}
\bibinfo{author}{S.~Xu}, \bibinfo{author}{Y.-m. Chen}, \bibinfo{author}{X.-y.
  Meng}, \bibinfo{author}{R.~Pan}, \bibinfo{author}{A.-x. Yan},
  \bibinfo{author}{Z.-m. Li}, \bibinfo{author}{Z.-l. Li},
\newblock \bibinfo{title}{Computational-assisted protein engineering to develop
  thermostable and highly active catalase for industrial and biocatalytic
  applications},
\newblock \bibinfo{journal}{Bioresource Technology} \bibinfo{volume}{437}
  (\bibinfo{year}{2025}) \bibinfo{pages}{133081}.
%Type = Article
\bibitem[{Johnson et~al.(2025)Johnson, Fu, Viknander, Goldin, Monaco,
  Zelezniak, and Yang}]{johnson2025protein_scoring}
\bibinfo{author}{S.~R. Johnson}, \bibinfo{author}{X.~Fu},
  \bibinfo{author}{S.~Viknander}, \bibinfo{author}{C.~Goldin},
  \bibinfo{author}{S.~Monaco}, \bibinfo{author}{A.~Zelezniak},
  \bibinfo{author}{K.~K. Yang},
\newblock \bibinfo{title}{Computational scoring and experimental evaluation of
  enzymes generated by neural networks},
\newblock \bibinfo{journal}{Nature Biotechnology} \bibinfo{volume}{43}
  (\bibinfo{year}{2025}) \bibinfo{pages}{396--405}.
%Type = Article
\bibitem[{Ding et~al.(2025)Ding, Zhang, Kong, Hess, and
  Zhang}]{ding2025denovo_cutinase}
\bibinfo{author}{Y.~Ding}, \bibinfo{author}{S.~Zhang},
  \bibinfo{author}{X.~Kong}, \bibinfo{author}{H.~Hess},
  \bibinfo{author}{Y.~Zhang},
\newblock \bibinfo{title}{Replicating pet hydrolytic activity by positioning
  active sites with smaller synthetic protein scaffolds},
\newblock \bibinfo{journal}{Advanced Science} \bibinfo{volume}{12}
  (\bibinfo{year}{2025}) \bibinfo{pages}{2500859}.
%Type = Article
\bibitem[{Seki et~al.(2025)Seki, Guo, Akpinaroglu, and
  Kortemme}]{seki2025Ephb1_proteinMPNN_frame2seq}
\bibinfo{author}{K.~Seki}, \bibinfo{author}{A.~B. Guo},
  \bibinfo{author}{D.~Akpinaroglu}, \bibinfo{author}{T.~Kortemme},
\newblock \bibinfo{title}{A combinatorial mutational map of active non-native
  protein kinases by deep learning guided sequence design},
\newblock \bibinfo{journal}{bioRxiv}  (\bibinfo{year}{2025}).
%Type = Article
\bibitem[{Yeh et~al.(2023)Yeh, Norn, Kipnis, Tischer, Pellock, Evans, Ma, Lee,
  Zhang, Anishchenko et~al.}]{yeh2023denovo_luciferase}
\bibinfo{author}{A.~H.-W. Yeh}, \bibinfo{author}{C.~Norn},
  \bibinfo{author}{Y.~Kipnis}, \bibinfo{author}{D.~Tischer},
  \bibinfo{author}{S.~J. Pellock}, \bibinfo{author}{D.~Evans},
  \bibinfo{author}{P.~Ma}, \bibinfo{author}{G.~R. Lee}, \bibinfo{author}{J.~Z.
  Zhang}, \bibinfo{author}{I.~Anishchenko}, et~al.,
\newblock \bibinfo{title}{De novo design of luciferases using deep learning},
\newblock \bibinfo{journal}{Nature} \bibinfo{volume}{614}
  (\bibinfo{year}{2023}) \bibinfo{pages}{774--780}.
%Type = Article
\bibitem[{Wang et~al.(2024)Wang, Chen, Ding, Cao, Qin, Niu, Zhuang, Li, Feng,
  Xu et~al.}]{wang2024Propend}
\bibinfo{author}{Z.~Wang}, \bibinfo{author}{B.~Chen},
  \bibinfo{author}{K.~Ding}, \bibinfo{author}{J.~Cao},
  \bibinfo{author}{M.~Qin}, \bibinfo{author}{Y.~Niu},
  \bibinfo{author}{X.~Zhuang}, \bibinfo{author}{X.~Li},
  \bibinfo{author}{K.~Feng}, \bibinfo{author}{T.~Xu}, et~al.,
\newblock \bibinfo{title}{Multi-purpose controllable protein generation via
  prompted language models},
\newblock \bibinfo{journal}{bioRxiv}  (\bibinfo{year}{2024})
  \bibinfo{pages}{2024--11}.
%Type = Article
\bibitem[{Tan et~al.(2024)Tan, Wang, Wu, Hong, and Zhou}]{tan2024ProtREM}
\bibinfo{author}{Y.~Tan}, \bibinfo{author}{R.~Wang}, \bibinfo{author}{B.~Wu},
  \bibinfo{author}{L.~Hong}, \bibinfo{author}{B.~Zhou},
\newblock \bibinfo{title}{Retrieval-enhanced mutation mastery: Augmenting
  zero-shot prediction of protein language model},
\newblock \bibinfo{journal}{arXiv preprint arXiv:2410.21127}
  (\bibinfo{year}{2024}).
%Type = Article
\bibitem[{Zimmerman et~al.(2024)Zimmerman, Alon, Levin, Koganitsky, Shpigel,
  Brestel, and Lapidoth}]{zimmerman2024CoSann_polyphosphate_glucokinase}
\bibinfo{author}{L.~Zimmerman}, \bibinfo{author}{N.~Alon},
  \bibinfo{author}{I.~Levin}, \bibinfo{author}{A.~Koganitsky},
  \bibinfo{author}{N.~Shpigel}, \bibinfo{author}{C.~Brestel},
  \bibinfo{author}{G.~D. Lapidoth},
\newblock \bibinfo{title}{Context-dependent design of induced-fit enzymes using
  deep learning generates well-expressed, thermally stable and active enzymes},
\newblock \bibinfo{journal}{Proceedings of the National Academy of Sciences}
  \bibinfo{volume}{121} (\bibinfo{year}{2024}) \bibinfo{pages}{e2313809121}.
%Type = Article
\bibitem[{Cheng et~al.(2024)Cheng, Mao, Tang, Yang, Cheng, Wang, Gu, Han, Chen,
  Li et~al.}]{cheng2024ProMEP_tnbp}
\bibinfo{author}{P.~Cheng}, \bibinfo{author}{C.~Mao},
  \bibinfo{author}{J.~Tang}, \bibinfo{author}{S.~Yang},
  \bibinfo{author}{Y.~Cheng}, \bibinfo{author}{W.~Wang},
  \bibinfo{author}{Q.~Gu}, \bibinfo{author}{W.~Han}, \bibinfo{author}{H.~Chen},
  \bibinfo{author}{S.~Li}, et~al.,
\newblock \bibinfo{title}{Zero-shot prediction of mutation effects with
  multimodal deep representation learning guides protein engineering},
\newblock \bibinfo{journal}{Cell Research} \bibinfo{volume}{34}
  (\bibinfo{year}{2024}) \bibinfo{pages}{630--647}.
%Type = Article
\bibitem[{Lambert et~al.(2026)Lambert, Tavakoli, Dharuman, Yang, Bhethanabotla,
  Kaur, Hill, Ramanathan, Anandkumar, and
  Arnold}]{lambert2026tryptophane_synthase}
\bibinfo{author}{T.~Lambert}, \bibinfo{author}{A.~Tavakoli},
  \bibinfo{author}{G.~Dharuman}, \bibinfo{author}{J.~Yang},
  \bibinfo{author}{V.~Bhethanabotla}, \bibinfo{author}{S.~Kaur},
  \bibinfo{author}{M.~Hill}, \bibinfo{author}{A.~Ramanathan},
  \bibinfo{author}{A.~Anandkumar}, \bibinfo{author}{F.~H. Arnold},
\newblock \bibinfo{title}{Sequence-based generative ai design of versatile
  tryptophan synthases},
\newblock \bibinfo{journal}{Nature Communications}  (\bibinfo{year}{2026}).
%Type = Article
\bibitem[{Li et~al.(2024)Li, Lou, Sun, and
  Li}]{li2024UDP_glucose_pyrophosphorylase}
\bibinfo{author}{Z.~Li}, \bibinfo{author}{M.~Lou}, \bibinfo{author}{C.~Sun},
  \bibinfo{author}{Z.~Li},
\newblock \bibinfo{title}{Engineering a robust udp-glucose pyrophosphorylase
  for enhanced biocatalytic synthesis via proteinmpnn and ancestral sequence
  reconstruction},
\newblock \bibinfo{journal}{Journal of Agricultural and Food Chemistry}
  \bibinfo{volume}{72} (\bibinfo{year}{2024}) \bibinfo{pages}{15284--15292}.

\end{thebibliography}

Papers of particular interest, published within the period of review, have been highlighted as:

* of special interest

** of outstanding interest

\end{document}